\documentclass[sigconf]{acmart}
\usepackage{booktabs}
\usepackage{multirow}
\usepackage{longtable}
\AtBeginDocument{%
  \providecommand\BibTeX{{%
    \normalfont B\kern-0.5em{\scshape i\kern-0.25em b}\kern-0.8em\TeX}}}

\copyrightyear{2024}
\acmYear{2024}
\setcopyright{acmlicensed}\acmConference[CHI '24]{Proceedings of the CHI Conference on Human Factors in Computing Systems}{May 11--16, 2024}{Honolulu, HI, USA}
\acmBooktitle{Proceedings of the CHI Conference on Human Factors in Computing Systems (CHI '24), May 11--16, 2024, Honolulu, HI, USA}
\acmDOI{10.1145/3613904.3642087}
\acmISBN{979-8-4007-0330-0/24/05}

\acmSubmissionID{198}

\begin{document}

\title{Establishing Heuristics for Improving the Usability of GUI Machine Learning Tools for Novice Users}


\author{Asma Yamani}
\authornotemark[1]
\orcid{0000-0002-6277-8972}
\affiliation{%
\department{Information and Computer Science}
  \institution{King Fahd University of Petroleum and Minerals}
  \city{Dhahran}
  \country{Saudi Arabia}
  \postcode{31261}
}
\email{201906630@kfupm.edu.sa}

\author{Haifa Alshammare}
\orcid{0009-0007-1522-1147}
\authornote{Both authors contributed equally to this research.}
\affiliation{%
\department{Information and Computer Science}
  \institution{King Fahd University of Petroleum and Minerals}
  \city{Dhahran}
  \country{Saudi Arabia}
  \postcode{31261}
}
\email{202110890@kfupm.edu.sa}

\author{Malak Baslyman}
\orcid{0000-0003-4002-4480}
\authornote{Corresponding author.}

\affiliation{%
\department{Information and Computer Science}
  \institution{King Fahd University of Petroleum and Minerals}
  \city{Dhahran}
  \country{Saudi Arabia}
  \postcode{31261}
  }
\email{malak.baslyman@kfupm.edu.sa}

\renewcommand{\shortauthors}{Yamani et al.}


\begin{abstract}

Machine learning (ML) tools with graphical user interfaces (GUI) are facing demand from novice users who do not have the background of their underlying concepts. These tools are frequently complex and pose unique challenges in terms of interaction and comprehension by novice users. There is yet to be an established set of usability heuristics to guide and assess GUI ML tool design. To address this gap, in this paper, we extend Nielsen's heuristics for evaluating GUI ML Tools through a set of empirical evaluations. To validate the proposed heuristics, user testing was conducted by novice users on a prototype that reflects those heuristics. Based on the results of the evaluations, our new heuristics set improves upon existing heuristics in the context of ML tools. It can serve as a resource for practitioners designing and evaluating these tools.\end{abstract}

\ccsdesc[500]{Computing methodologies~Machine learning}
\ccsdesc[500]{Human-centered computing~HCI design and evaluation methods}
\ccsdesc[500]{Software and its engineering~Software usability}

\keywords{Usable, ML tools, GUI machine learning tools, Weka, Knime, Heuristic evaluation (HE), Cognitive walkthrough (CW), Novice users, User testing, SUS}


\maketitle
\section{Introduction}
\label{intro}
Machine learning (ML) is a subset of artificial intelligence that uses computerized methods to solve issues based on historical data and knowledge without requiring unnecessary changes to the core process\cite{Sandhu2018}. With the sheer of data available to analyze and build accurate prediction models, machine learning tools have become necessary to add to a professional's toolkit in various fields. Thus, machine learning has gained immense attraction in recent years. Several software tools have been developed to make machine learning programming easier and faster. These software tools come in the form of programming languages, libraries/packages, graphical user interface-based programs, platforms, toolkits, and combinations of any of these~\cite{ogunleye2019development}. GUI  ML tools, such as Weka, KNIME, and RapidMiner, are easier to use, to some extent, for novice users to use rather than code. According to Dudley and Kristensson~\cite{Dudley2018}, machine learning techniques are steadily entering novice users' lives. Thus, enabling individuals to interact with such technologies efficiently will be a significant design problem. This pushed machine learning from being an underlying technological infrastructure unaddressed by HCI researchers to the fore of interface design and user experience~\cite{Gillies2016}. From this perspective, new research practices and design methods are required to address the end-user interaction challenges with ML tools~\cite{Burrell2016,Dove2017}. However, while there is growing acknowledgment of the issues that user experience (UX) and visualization practitioners encounter when working with machine learning~\cite{Dove2017,yang2018mapping,Yang2018,Yang2018b}; the usability of current ML tools has not been studied adequately using a formal process as previous studies are either old and the tools they have studied are outdated or updated~\cite{chen2007survey,boscarioli2014analyzing} or conducted informal evaluations that had key information of the study design missing~\cite{ogunleye2019development,Bansal2018ToolsUI}. Subsequently, the heuristic evaluation method still needs to be updated to address the challenges of ML tools to have a better end-user experience.

Although Nielsen heuristics, the most well-known heuristics throughout HCI~\cite{Nielsen1990}, can be used to evaluate GUI ML tools and was able to identify usability issues in GUI ML tools in the first part of our study, several studies have suggested that Nielsen's heuristics are general and do not address domain-specific issues ~\cite{mankoff2003heuristic,murad2019revolution,wei2018evaluating,langevin2021heuristic}. The need to extend Nielsen's heuristics to accommodate GUI ML tools for novice users arises since these tools are actually an interface to underlying complex concepts that novice users are not familiar with. This increases the burden on the tool from performing certain tasks to helping the user navigate complex concepts and gain competence in the field over time. Hence, in this work, we sought to extend Nielsen's heuristics for ML tools to enhance their design and usage concerning novice users. Therefore, we pose the following research question:
\begin{itemize}
    \item \textbf{RQ1: To what degree are the current GUI ML tools usable?}
    \item \textbf{RQ2: What are the suitable heuristics for evaluating GUI ML tools usability?}
\end{itemize}
To address the first question, we conducted a cognitive walkthrough and heuristic evaluation for two commonly used GUI ML tools (Weka and KNIME). For the second question, we empirically propose 14 heuristics for GUI ML tools by applying a four-phase design process, including heuristics generation, prototyping, validation, and revising. Based on the results, these new heuristics are more effective in enhancing the usability of GUI ML tools. The contribution of this work is as follows: (1) Empirical usability evaluation of Weka and KNIME GUI ML tools; (2) A set of verified heuristics that practitioners and researchers may use to evaluate GUI ML tools formally. The rest of the paper is structured as follows: Section~\ref{LR} provides a literature review of the related work on GUI ML tools usability evaluations and heuristics. Section~\ref{meth} emphasizes the fundamental details of this research methodology. The details of establishing the new heuristics stages, as well as their experimental design, results, and analysis, are presented in Sections~\ref{Ph1},\ref{Ph2},\ref{Ph3}, and~\ref{Ph4}. Sections~\ref{Dis} go through the discussion and limitations of the study. The conclusion and future work are covered in Section~\ref{conc}.


\section{Related Work}
\label{LR}
\subsection{Usability of GUI ML tools}
Though the importance of having a highly usable GUI ML tool to foster and encourage adaptations of ML from all sectors, the usability of GUI ML tools or tools in related domains, such as data mining, has not been addressed adequately in the literature. One of the earliest studies conducted to address the usability evaluation of GUI ML tools dated back to 2007 by X. Chen et al.~\cite{chen2007survey}. It covered 12 open-source data mining tools. The authors evaluated the usability of tools based on human interaction, interoperability, and expandability, among other aspects such as functionality and data source. In this study, KNIME and AlphaMiner had the highest scores regarding usability. Seven years later, a study investigated the usability of five data mining tools: KNIME, Orange Canvas, RapidMiner, and Weka. The study conducted a cognitive walkthrough to perform the evaluation on students and lecturers from the CS department. The results concluded that lecturers found Weka was easier to use than KNIME, while students thought the opposite. Both students and lecturers agreed that in many aspects of the evaluation, Weka and KNIME were the best amongst the other GUI data mining tools~\cite{boscarioli2014analyzing}.
As for more recent studies, in 2018, a study attempted to compare the data analytics tools Python, SPSS, R, SAS, and Weka concerning multiple factors, including usability. The study gave all the tools 4 out of 5 when it came to usability, except for R, which got 4.5~\cite{Bansal2018ToolsUI}. Another recent study in 2019 reported the SUS evaluation of GUI ML tools: Orange Canvas, RapidMiner, Weka, and KNIME against a tool the study authors developed~\cite{ogunleye2019development}. Aside from the developed software, Weka had a higher satisfaction rate than RapidMiner and KNIME. \par

Based on the surveyed literature regarding GUI ML tools, the early studies contain tools that are now obsolete or updated~\cite{chen2007survey,boscarioli2014analyzing}. Also, some studies based their evaluation on the authors' or participants' perspectives without following any formal definition of usability or following any of the formal usability evaluation methods~\cite{chen2007survey}. As for other studies~\cite{Bansal2018ToolsUI,ogunleye2019development}, some important experimental setup details were missing, such as the number of participants, the number of questions, or the tasks they completed before performing the evaluation, hence a reliability issue. Although~\cite{boscarioli2014analyzing} reported using cognitive walk; the approach was not fully implemented as it was conducted on the task level, not on the action level. In addition, the survey questions were given at the end of the evaluation, not task by task. The question also did not align with the four questions from~\cite{wharton1994cognitive}. Moreover, in~\cite{boscarioli2014analyzing}, the scenario investigated was extremely limited to four tasks only, which is not a representative set of tasks considering all the steps of the ML development life cycle. Therefore, although the results may be valuable for the data mining community, they may not be generalized to the issues faced when developing and deploying an ML model.

\subsection{Interactive Machine Learning}
Another paradigm for designing ML tools for novice users is Interactive Machine Learning (iML)~\cite{IML}. IML is an interaction paradigm in which a user or user group iteratively builds and refines a mathematical model to describe a concept through iterative cycles of input and review~\cite{dudley2018review}. Current iML tools focus on building a highly accessible tool that provides the basic functionality for a specific problem type that novice users need and it can also be used for educational purposes. Some problems that are currently addressed are images, voices, and pose classification in Teachable Machine~\cite{Carney2020}, regression in  TREX~\cite{iii2021explore}, and transfer learning in images~\cite{Mishra2021}.

The scaffolding approach followed in iML tools allowed the users to focus their mental efforts on integrating their domain knowledge into the models. For example, Teachable Machine~\cite{Carney2020} was recently evaluated in the context of aiding the development of prototypes by UX practitioners (UXP) to propose a proof-of-concept design for a new ML-enabled application~\cite{Feng2023}. Participants found it valuable to interactively explore different combinations of model classes and visualize the data. Although not all were able to interpret the probability scores correctly, coming from no background in ML, it helped in design choices and in identifying limitations and possible biases. The ML module integrated with the app prototype was deemed helpful in communicating with stakeholders and managers~\cite{10.1145/3563657.3596101}
\par
A limitation in current iML tools raised in ~\cite{Mishra2021} was that overuse of scaffolding which comes with the risk of forcing individuals to follow a pattern conveyed by the interface. Another limitation observed is the narrow purpose and functionality of each of the tool as it came as a trade-off with usability. Also there are no heuristics or guidelines to reproduce such usable tools. Hence, in this work, we aim to evaluate the current most popular GUI ML tools for usability following formal usability evaluation approaches in order to develop heuristics that can be used by designers to develop usable ML tool.

\subsection{The role of AI in GUI generation}
Generative AI is starting to play a significant role in generating prototypes and GUIs. Akin is a  UI wireframe generator built using a fine-tuned Self-Attention Generative Adversarial Network trained with the semantic representations of UI images of 500 UI wireframes from different design patterns. Human evaluation of the resulting wireframes against designers' wireframes shows similar quality. In 50\% of the occasions, designers identified Akin-generated wireframes as designer-made 50\% of the time~\cite{10.1145/3397482.3450727}. In WireGen~\cite{feng2023designing}, the researchers leveraged the contextual knowledge in LLMs. They fine-tuned a gpt-3.5-turbo with 1000 user interfaces to generate mid-fidelity wireframes that go beyond the shape and text placeholders of low-fidelity wireframes to include real content, semantic icons, and interactive elements. Aside from the active research in the area, several companies are emerging on this premise, including Semanttic~\footnote{https://www.semanttic.com} that can be used to regenerate and iterate over previously built interfaces using AI. There is also Builder.io\footnote{https://www.builder.io/blog/ai-figma}, a Figma plug-in that generates UI based on a text prompt; a similar service is provided by Galileo AI\footnote{https://www.usegalileo.ai/explore}. Dice\footnote{https://www.dice.design} creates UX strategies and suggests design inspiration based on the project description through a text prompt. With the progression of AI, it could result in enhancing the user experiance and create personalized UI/UX experiences per context.

\section{Research Methodology}
\label{meth}
\begin{figure*}[ht]
  \centering
  \includegraphics[width=0.9\textwidth]{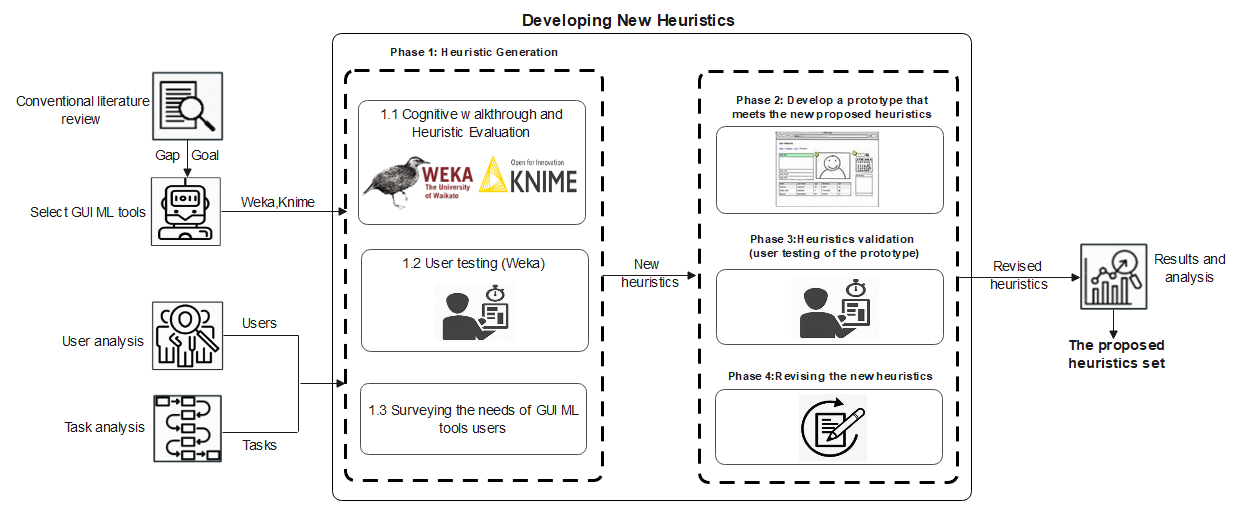}
    \Description[Research Methodology]{"A diagram shows the research methodology consisting of the following steps: conventional literature review followed by select GUI ML tools, user analysis, and task analysis. The output of these steps is used as an input to develop new heuristics. This process consists of 4 phases. The first phase is heuristic generation, which consists of 3 steps (cognitive walkthrough and heuristics evaluations, user testing of Weka, and surveying the needs of GUI ML tools). The second, third, and fourth phases are developing a prototype that meets the new proposed heuristics, validating the developed heuristics, and revising the heuristics, respectively. Then, the revised heuristics are input to results and analysis, and the output is the proposed heuristics phase" }

  \caption{Research Methodology\label{methology}}
  
\end{figure*}

To address our research questions, the following goal was formulated according to the Goal Question-Metric (GQM) approach~\cite{vanSolingen2002}: develop new heuristics for the purpose of enhancing the usability of GUI ML tools with respect to their effectiveness from the point of view of the researcher in the context of Nielsen’s ten heuristics, and novice users using Weka, and KNIME ML tools. Figure~\ref{methology} outlines the methodology we followed to achieve this goal. To address RQ1, we conducted a literature review to identify the most commonly used ML tools in addition to the goals and gaps (as discussed in Section~\ref{tool selection}). We also performed a user and task analysis to evaluate the usability of the selected GUI ML tools (as presented in Sections ~\ref{user analysis} and~\ref{task analysis}). To address RQ2, we followed four phases: (1) heuristics generation, (2) developing a prototype that meets the new proposed heuristics, (3) validation of the developed heuristics, and (4) revising the heuristics. We based Phases 1, 3, and 4 on~\cite{langevin2021heuristic,amershi2019guidelines}, but made some modifications. The following Sections (\ref{Ph1}–\ref{Ph4}) describe each phase in detail.

\subsection{GUI ML tools selection.}
\label{tool selection}
Among multiple GUI ML tools Weka~\cite{witten2002data} and KNIME~\cite{Berthold:2009:KKI:1656274.1656280} were chosen for multiple reasons:
\begin{enumerate}
    \item They are both widely utilized for data processing~\cite{dwivedi2016comprehensive,hussain2018educational,ratra2020experimental}, and they are among the top 10 machine learning tools for 2021~\cite{chowdhury2021}.
    \item KNIME is recommended for individuals who are novices at using such software~\cite{dwivedi2016comprehensive}. It is very robust with built-in components and extra functionality that can be brought from third-party libraries\cite{dwivedi2016comprehensive}. It also comes with a drag-and-drop interface. Based on the analysis, Weka comes second to KNIME because of its many built-in capabilities that do not require any programming or coding experience~\cite{dwivedi2016comprehensive,analytics2022}.
    \item  Weka and KNIME are both open-source and cross-platform software that appeals to all novice users.
    \item  Weka and KNIME are two of the common benchmarking tools\cite{alsalem2018systematic}, and to the best of our knowledge, no work evaluated their usability through heuristic evaluation or User testing usability methods.
    
\end{enumerate}

It is worth noting that both Weka and KNIME were used in the first two evaluation methods (Heuristic Evaluation and Cognitive Walkthrough). Only the tool with the least number of issues was chosen afterward to perform user testing evaluation and SUS evaluation to avoid frustration during user testing.\par
\subsection{Identifying targeted users (User Analysis).}
\label{user analysis}
In the era of data science, the knowledge of implementing the basic algorithms and methods related to machine learning and data analytics is essential for beginners in this field (novice users). Thus, we evaluated the usability of Weka, KNIME, and the developed prototype from the point of view of novice users to help in designing and developing usable ML tools based on the resulting usability issues and suggested recommendations. Fig.\ref{persona} (in the appendix) shows the persona of our ML tools novice user. \par

\subsection{Identifying user goal and Task analysis.}
\label{task analysis}
\begin{figure*}[ht]
  \centering
\Description[The nine stages of the machine learning workflow]{The nine stages of the machine learning workflow: model requirements, data collection, data cleaning, data labeling, feature engineering, model training, model evaluation, model deployment, and model monitoring.}

  \includegraphics[width=0.8\textwidth]{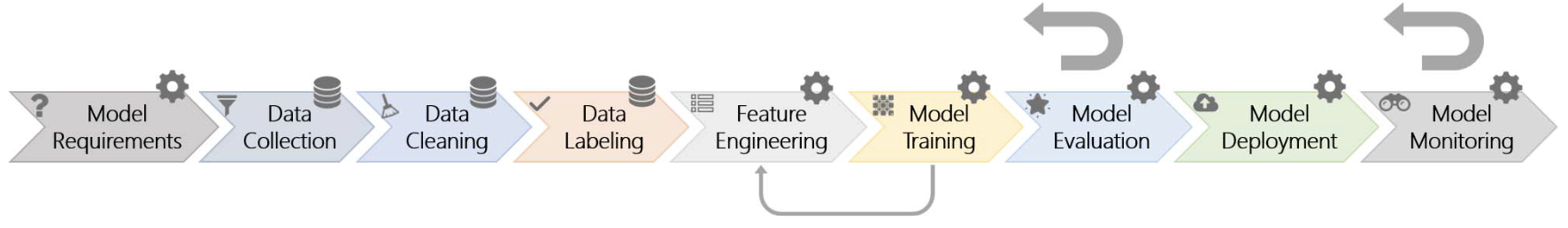}
  \caption{The nine stages of the machine learning workflow~\cite{amershi2019software}. \label{nine}}
  
\end{figure*}

\begin{figure*}[ht]
  \centering
  \Description[Task Analysis]{A diagram illustrating the task Analysis of predicting a price of a house goal, which consists of all sub-tasks that need to be performed sequentially to satisfy the goal.}

  \includegraphics[width=\textwidth]{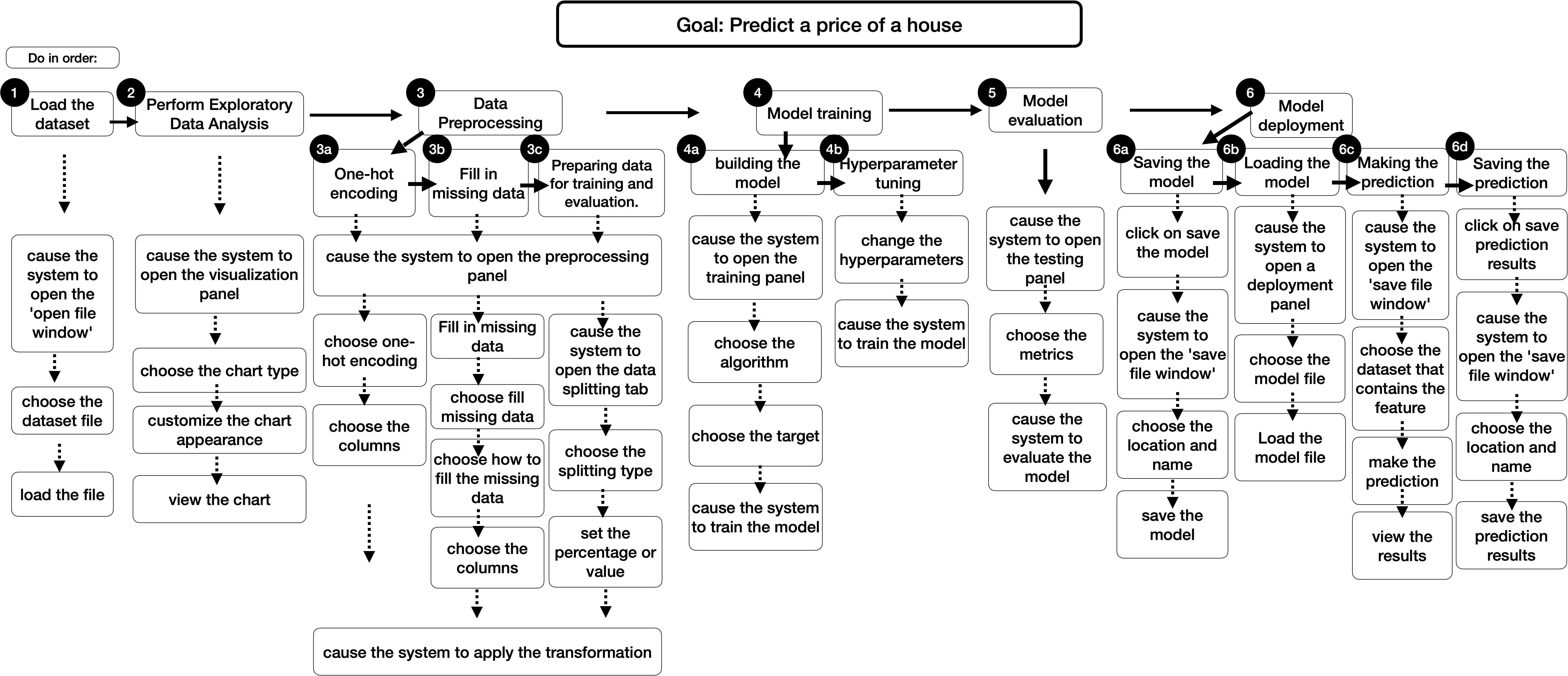}
  \caption{Task Analysis\label{task}}
  
\end{figure*}

To conduct the usability evaluations of the GUI ML tools, we identified the goal of using ML tools by novice users through a scenario that simulated a real-life case. From this goal, we identified and analyzed the tasks that will be used during usability evaluations (Heuristic evaluation and cognitive walkthrough) to help measure the usability metrics, such as success rate and errors and finding related issues.
\newline
\emph{Scenario:} Tiana is a real estate agent in California. She is trying to estimate the best prices for houses in multiple regions. Tiana's friend provided her with a dataset containing 20,000 records in an Excel sheet. The data consisted of the previous sales made by a real estate agent, along with a description of each house. Her friend suggested using machine learning to help her predict outcomes based on the features of the houses. Tiana had never worked with ML or ML tools before, so she found a tutorial and decided to follow it. As she was intimidated by programming Python, she found two cross-platform free ML GUI tools and decided to try them out.\newline
\emph{Goal:} Predict the price of a house based on a set of features with respect to historical data.\newline
\emph{Task Analysis:} The Tasks to achieve this goal follows, with some customization, the nine stages of the machine learning workflow~\cite{amershi2019software}, illustrated in Fig.\ref{nine}. We assume the requirements are clear and start by (1) Loading the labled data, and (2) performing exploratory data analysis have a quick look on the features. Then, some necessary data preprocessing including (3a) one-hot encoding that changes categorical data to numerical, (3b) filling in missing data, and (3c) splitting the data for training and testing. No feature engineering is performed since the user is a novice. Then, the model training starts, including (4a) building the model and (4b) performing hyperparameter tuning and building in an iterative manner. Following that is (5) model evaluation based on the appropriate metrics. Finally, the model deployment step occurs in which the user (6a) saves the model, then the same user or a different one (6b) load the model in inference mode, (6c) makes predictions on an independent dataset, and (6d) saves the prediction results. Model monitoring, from Fig.\ref{nine}, is also left out as it not a task novice users would do. The steps under each task are illustrated in Fig.\ref{task}. Exact wording of the tasks presented to the participants when performing user testing are in Table~\ref{tab: user-testing-q} (in the appendix) The Tasks are to be completed sequentially.\par

\section{Phase 1: Heuristics Generation}
\label{Ph1}
The first phase of this study was to generate the heuristics after modifying Nielsen's heuristics and adapting them to GUI ML tools based on empirically evaluating two GUI ML tools. 

\subsection{Experimental Design}
\label{Ph1-Experimental}
\subsubsection{Heuristic evaluation.} As a result of reviewing the literature, there were no heuristics proposed specifically for evaluating the ML tools. Accordingly, Nielsen's Ten usability heuristics~\cite{Nielsen1990}, shown in Table~\ref{tab:heuristics-11}, were used in this heuristic evaluation, which is the most common set of heuristics utilized. In this heuristic evaluation, all the UIs visited for performing each task presented in Fig.~\ref{task} were evaluated against the ten heuristics, looking for usability issues in the Weka and KNIME ML tool design. After that, a severity was assigned for all the issues found using Jakob Nielsen's severity rating scale~\cite{nielsen2022b}, in Table~\ref{tab: severity} (in the appendix). These severities help rank each issue's significance to the users using those ML tools. \par

\subsubsection{Cognitive walkthrough evaluation.}
A cognitive walkthrough aims to evaluate the ease of learning a system by exploration through the following four questions~\cite{wharton1994cognitive}:
\begin{enumerate}
    \item Will the user try to achieve the right effect?
    \item Will the user notice that the correct action is available?
    \item Will the user associate the correct action with the effect that the user is trying to achieve?
    \item If the correct action is performed, will the user see that progress is being made toward the solution of the task?
\end{enumerate}
\subsubsection{Surveying the needs of GUI ML tools users.}
\label{questionnaire}
An online questionnaire\footnote{\url{https://forms.gle/hhHgWXhNFbGQ4fW28}} that consists of twelve items arranged in three parts was used. The first part indicates three items about the experience with ML tools that the respondents have. The second and third parts show four items based on the perception of difficulties while using ML tools and five items based on characteristics(needs) that lead to more usable tools, respectively. Seven items are close-ended. Three of them are scored using a 5-point Likert scale from (1) `strongly disagree' to (5) `strongly agree', while four are multiple choices, and the rest five items are open-ended. After designing the questionnaire, it was tested by an expert to increase its validity and distributed through several social media channels to the ML community, targeting people who had previous experience with GUI ML tools. {This questionnaire is the only method where participants did have prior experience with GUI ML tools, we focused in the analysis on users with less than 6 months of experience and used the responses of participants with more than 6 months of experience to emphasize the novice users’ responses.} \par 

\subsubsection{User testing evaluation of Weka.} Usability testing is one of the most common methods to evaluate product usability by observing users while they accomplish Tasks with that product~\cite{roy2014some}. As known, it is crucial to test with users to get more usability issues, which could help in developing more comprehensive heuristics to solve them. Thus, we conducted an online moderated user testing of Weka since it is one of the most known GUI ML tools, as mentioned before, and had the least issues when it comes to the cognitive walkthrough and heuristic evaluation. According to Nielsen and Landauer, five users are enough to identify 85\% of the issues on most websites~\cite{liu2008usability}. Domas and Redish also supported this argument, claiming that five to twelve participants are sufficient for a test ~\cite{liu2008usability}. Therefore, in this evaluation, seven novice participants have been recruited. The evaluation was conducted in separate sessions for each participant, where the sessions were video recorded for analysis purposes(for identifying the errors type and description).
Each participant signed the consent form  and performed twelve Tasks presented in Table~\ref{tab: user-testing-q} in the appendix. While the participant was performing each task, the evaluator observed and recorded the number of errors made when deviated from the optimal way to perform the task, the time it takes to perform the task in seconds, and the task success with three success levels proposed by Nielsen (S = Success, F = Failure, P = Partial Success). The task success rate was computed as follows\cite{norman2001}: 
\begin{equation}
    success\_rate = (S + (P*0.5)/\textsc{No. of attempts}),
\end{equation}
where the \textsc{No. of attempts} is the number of overall attempts to perform the task, which is equal to the number of participants in the evaluation. These measures will compute the Task completion rate, time spent on each task, error rate, and error types for the analysis. After performing all Tasks, the participants were asked to fill up the System Usability Scale (SUS) to indicate their level of satisfaction with the tool as described in Section~\ref{SUSsection} below. Also, an informal interview was conducted with each participant to get some extra views and opinions.\par

\subsubsection{The System Usability Scale (SUS).} 
\label{SUSsection}
The System Usability Scale (SUS) was used to assess the level of users' perceived usability. It is a reliable tool for evaluating the user's satisfaction level with a system, represented as a questionnaire of 10 items with 5-point Likert scale ratings from strongly disagree (1) to strongly agree (5)~\cite{bindu2015secure}. This SUS score will be calculated as follows\cite{bindu2015secure}:
\begin{enumerate}
    \item Subtract one from the score of each odd-numbered question.
    \item Subtract the score of the even-numbered question from 5.
    \item Sum all the generated values (from all questions) and then multiply the total by 2.5 to obtain the overall SUS score.
\end{enumerate}
The SUS score can be transformed into aspects of Adjectives, Grade, and Acceptable\cite{sauro2022}. Acceptability ranges are used to interpret the SUS score depending on user satisfaction as Not Acceptable, Marginal, and Acceptable, as shown in Table 5~\cite{sauro2022}. On the Grade scale, SUS scores are classified as A, B, C, D, E, and F grades from superior performance to failing performance, with C as an average~\cite{sauro2022}. Moreover, the Adjective rating is an adjective that converts the SUS numerical score into an absolute usability rating, which includes Worst Imaginable, Awful, Poor, OK, Good, Excellent, and Best Imaginable~\cite{bangor2009determining}. Two open-ended questions were also included about what the user liked most about the tool and what the user preferred to change.  
\subsection{Results and Analysis}
\label{Ph1-results}
\subsubsection{Heuristic Evaluation.}
\begin{figure}[ht]
  \centering
  \includegraphics[width=0.9\linewidth]{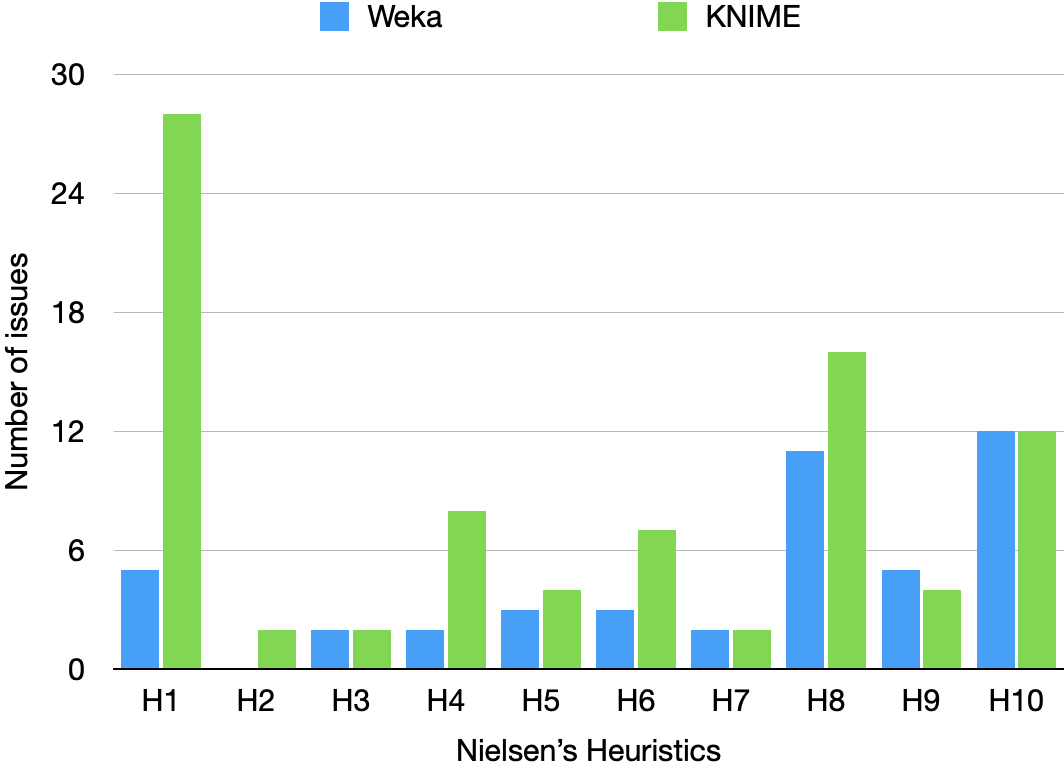}
  \caption{Issues discovered during the Heuristic Evaluation heuristic number. \label{heu}}
    \Description[Issues discovered during the Heuristic Evaluation heuristic number.]{A bar chart displaying the number of issues discovered during the Heuristic Evaluation of WEKA and KINME regarding the 10 Nielsen heuristics. For WEKA, 5,0,2,2,3,3,2,11,5, and 12 issues for the heuristics number 1,2,3,4,5,6,7,8,9, and 10, respectively. For KNIME, 28,2,2,8,4,7,2,16,4 and 12 issues for the heuristics number 1,2,3,4,5,6,7,8,9, and 10, respectively}

\end{figure}

Heuristic Evaluation was conducted based on Nielsen's Heuristics, in Table~\ref{tab:heuristics-11}, by two evaluators. The evaluation revealed 45 issues for Weka, and 85 issues with KNIME, which is 1.8x the issues with Weka. Around 60\% of the issues were categorized as major usability problems, 15\% as minor usability problems. While Weka sustained minimal catastrophic usability problems, KNIME had 14 catastrophic usability problems. The distribution of the issues based on severity across tools is illustrated in Fig.~\ref{fig:sev} (in the appendix). An issue that was recognized across tasks and both tools was the lack of adequate help and documentation, as there was no proper offline help or documentation, and even the online help was hard to navigate. Another heuristic that was violated multiple times was the first heuristic related to 
the visibility of system status, as the systems lacked proper feedback in multiple instances. Also, the $8^{th}$ heuristic, which is related to Aesthetic and minimalist design, as too much unnecessary information and options are present at the same time for both tools. The distribution of the issues based on heuristic numbers across tools is illustrated in Fig.\ref{heu}. Also, the issues, along with the violated heuristic number, the severity, and the associated tasks, are present in Table~\ref{tab:HE-results} (in the appendix).

\begin{table*}
\caption{Nielsen’s Ten Heuristics.}
\label{tab:heuristics-11}
\begin{tabular}{@{}ll@{}}
\toprule

\multicolumn{1}{c}{\textbf{No}} & \multicolumn{1}{c}{\textbf{Nielsen’s Heuristics}} \\ \midrule
H1                                                   & Visibility of system status.                                                      \\
H2                                                  & Match between system and the real world.                                                            \\
H3                                                 & User control and freedom.                                                          \\ H4                                                  & Consistency and standards.                                                            \\ H5                                                  & Error prevention.                                                            \\ H6                                                  & Recognition rather than recall.                                                            \\ H7                                                  & Flexibility and efficiency of use.                                                            \\ H8                                                  & Aesthetic and minimalist design.                                                            \\ H9                                                  & Help users recognize, diagnose, and recover from errors.                                                            \\ H10                                                  & Help and documentation.                                                            \\ \bottomrule
\end{tabular}%

\end{table*}

\subsubsection{Cognitive walkthrough.}
The cognitive walkthrough revealed 12 issues related to the usability of Weka and KNIME in 42 instances out of which 30 instances belonged to KNIME, illustrated in Fig.~\ref{CWR}(in the appendix). The user in the scenario followed a tutorial associated with reaching the goal using Python. The reason for choosing a tutorial based on Python, not GUI ML tools, is that it mimics the real-life case as it is the type of tutorial found on the web. This presents an extra challenge in our cognitive walkthrough evaluation. The  Despite this, five issues are related to question 2, where many actions were not visible to the user. Five issues are related to question 3, where the step is there on the screen, but the user might have some difficulty associating. Finally, two issues are discovered related to not giving proper feedback when performing an action. In Table~\ref{tab:CWR}(in the appendix), we enumerate the issues found, along with the Action item they were discovered in, along with the suggestions.\newline


\begin{figure}[t]
  \centering
  \includegraphics[width=0.9\linewidth]{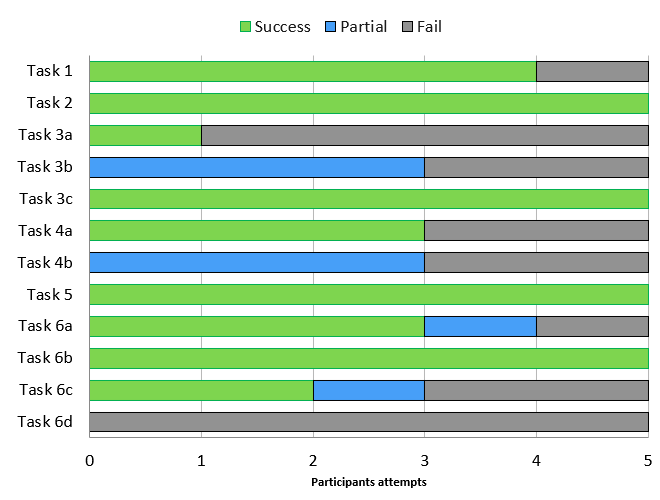}
  \caption{Success Rate Ratio\label{Success-Rate-Ratio-p1}}
  \Description[Success Rate Ratio.]{A bar chart displaying the Success Rate Ratio of the 12 tasks performed by five participants during the user testing. The Success Rate Ratio of Task 1 is success for participants No. 1, 2, 3, and 4 and fail for participants No. 5. The Success Rate Ratio of Task 2 is success for all participants. The Success Rate Ratio of Task 3 is success for Participant No. 1 and fail for Participants No. 2,3,4 and 5. The Success Rate Ratio of Task 4 is partially success for participants No. 1, 2, and 3, and fail for participants No. 4 and 5. The Success Rate Ratio of Task 5 is success for all participants. The Success Rate Ratio of Task 6 is success for participants No. 1, 2, and 3 and fail for participants No. 4 and 5. The Success Rate Ratio of Task 6 is success for all participants. The Success Rate Ratio of Task 7 is partially success for participants No. 1, 2, and 3, and fail for participants No. 4 and 5. The Success Rate Ratio of Task 8 is success for all participants. The Success Rate Ratio of Task 9 is success for participants No. 1, 2, and 3, partially success for Participant No. 4, and fail for Participant No. 5. The Success Rate Ratio of Task 10 is success for all participants. The Success Rate Ratio of Task 11 is success for participants No. 1 and 2, partially success for participants No. 3, and fail for participants No. 4 and 5. The Success Rate Ratio of Task 12 fails for all participants.}

\end{figure}
\subsubsection{User testing evaluation for the Weka.}
In this type of evaluation, we aimed to measure the proficiency and error rate when performing the set of 12 tasks that a novice user will most likely encounter when using ML tools. We initially recruited (N = 7) novice participants to the evaluation study, but (N = 2) had to withdraw since Weka on their devices did not recognize the file used for training without showing any error messages.\par
For measuring the \textit{proficiency}, two metrics were used: The success rate as suggested by Jakob Nielsen ~\cite{norman2001} and the completion time. The success rate per task is presented in Table~\ref{SucesssRatePhase1}, while the details of the attempts are illustrated in the stacked bar chart presented in Fig.\ref{Success-Rate-Ratio-p1}. In addition to the success rate for proficiency, we measured the average \textit{completion time}. The Tasks that were executed successfully across participants were: (1)Task 2 Performing Exploratory Data analysis represented in visualizing the correlations between the different attributes (2) Task 3c: Preparing the data for training and testing by splitting the data into training and test sets. (3)Task 5: Model evaluation represented in reporting the mean absolute error. (4)Task 6b: loading the model. It was notable that Tasks 2, 3c, and 5 were among the least Tasks to violate Nielsen's heuristics during the heuristic evaluation or present issues during the cognitive walkthrough. As for Task 6b, it had a reciprocal Task before it (Task 6a: Saving the model), in which the participants were able to learn how to deal with Task 6b. \par
With an average completion time of 48 minutes for the complete session, the average completion time per Task is 4 minutes. Generally, tasks with higher success rates have lower completion time, illustrated in Fig.\ref{fig:avgp1} (in the appendix).

\begin{table}
\caption{Average success rate across Tasks \label{SucesssRatePhase1}
}
\begin{tabular}{@{}llll@{}}
\toprule
\textbf{Task ID} & \textbf{Success rate} & \textbf{Task ID} & \textbf{Success rate} \\ \midrule
\textbf{Task 1}                          & \underline{\textbf{0.8}}                            & \textbf{Task {4b}}  & 0.3                                           \\
\textbf{Task 2}                          & \underline{\textbf{1.0}}                            & \textbf{Task {5}}  & \underline{\textbf{1.0}}                            \\
\textbf{Task {3a}}                          & 0.2                                           & \textbf{Task {6a}}  & \underline{0.7}                                     \\
\textbf{Task {3b}}                          & 0.3                                           & \textbf{Task {6b}} & \underline{\textbf{1.0}}                            \\
\textbf{Task {3c}}                          & \underline{\textbf{1.0}}                            & \textbf{Task {6c}} & \underline{0.5}                                     \\
\textbf{Task {4a}}                          & \underline{0.6}                                     & \textbf{Task {6d}} & 0.0                                           \\ \midrule
\textbf{Average}                         & \multicolumn{3}{c}{0.525}                                                                                                                \\ \bottomrule
\end{tabular}
\end{table}
As for Error handling, we explored the error rate by task, and then classified whether the error was the result of a user accident or caused by confusion. Fig.\ref{Error-rate-p1}(in the appendix), shows a comparison between the error rates of different Tasks. As for the source of the error, all errors were of \emph{error by confusion class}. The top source of confusion, confirmed to what was previously found by the heuristic evaluation cognitive walkthrough, was the lack of clear labels or buttons. In addition, with the deep nested lists, the participants could not find what they were looking for, which led to moving from the correct path they were initially on.

\subsubsection{SUS of Weka.}
\begin{figure}[ht]
  \centering
  \includegraphics[width=0.9\linewidth]{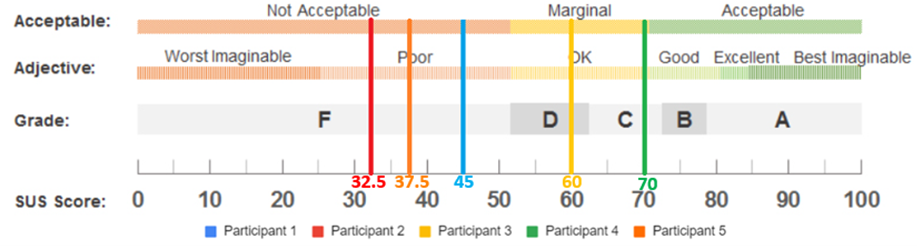}
  \caption{SUS evaluation of Weka.}
        \Description[SUS evaluation of Weka.]{
 A diagram showing the SUS scores of five participants with WEKA. The SUS scores of participants 1, 2, 3, 4, and 5 are 45, 32.5, 60, 70, and 37.5, respectively.}

  \label{sus-p1}
\end{figure}

An online SUS questionnaire was filled out by five participants who had been exposed to the Weka tool for the first time. The SUS score was calculated as discussed in section~\ref{Ph1-Experimental} using the equation from~\cite{suharsih2021usability}.
 Fig.\ref{sus-p1}, illustrates Weka satisfaction measures regarding the SUS score interpretation of all participants. The results of SUS scores were interpreted into different categories: Adjectives, Grade, and Acceptable. As shown in Fig.\ref{sus-p1}, three participants (1,2, and 5) revealed that Weka is not acceptable, with 'poor' usability and failing performance with an 'F' grade. On the other hand, Weka scored a marginal satisfaction level with 'OK' usability for the other two participants (3 and 4) with 'C' and 'D' grades, which indicates that Weka has an average level of satisfaction. On average, Weka has a 49 SUS score among all the participants, so we can conclude that it is considered not acceptable based on Table~\ref{tab:sus-range}. 
As for the open-ended question responses, a participant thought that the hint box was helpful, while another participant commented that the hint box was too long and that it should be concise. Another participant suggested improving the associated labels of the buttons and tabs. Also, using buttons instead of right-clicks as they were not being used in an intuitive manner in some places. Finally, to guide the process, one participant suggested that the upper bar should be about the main processes in machine learning.

\begin{table}
\caption{Acceptability ranges of SUS score}
\label{tab:sus-range}
\begin{tabular}{@{}ll@{}}
\toprule

\multicolumn{1}{c}{\textbf{SUS Score}} & \multicolumn{1}{c}{\textbf{Interpretation}} \\ \midrule
Less than 50                                                   & Not Acceptable                                                      \\
From 50 to 70                                                  & Marginal                                                            \\
Grater than 70                                                 & Acceptable                                                          \\ \bottomrule
\end{tabular}%

\end{table}
\subsection{Surveying the needs of GUI ML tools users}
An online questionnaire described in Section~\ref{questionnaire} was the method used for collecting qualitative or quantitative data about the users' needs from GUI ML tools. The questionnaire was distributed through several social media channels to those who have an experience in ML tools. {In this method we focus on the novice users needs and we corroborate on them using the expert users' opinions} Descriptive statistics were used to analyze the collected responses, as shown below.
\subsubsection{Descriptive Analysis of Responses.}
The total number of valid responses was 27. The majority of them have used Weka, as shown in Fig.\ref{Fig7} (in the appendix), while only two have used AZURE and RapidMiner tools, and nothing for KNIME. Regarding the experience, $52\%$  are novice users with experience of fewer than six months, and others are considered experts. Additionally, $56\%$ of the respondents have taken 1-3 courses in the field, and $52\%$ of them are Weka users.

In terms of perception of difficulties while using ML tools, eight respondents ($29.6\%$) disagree that the interaction with ML tools is straightforward and apparent; the bulk of them ($63\%$) have attended 1-3 courses in the field, while $50\%$ are novice users. Seven respondents explained why they thought the interaction with ML tools was not understandable and clear; novice users were concerned about guidance and learnability, while experts were concerned about trust and user interface issues.

\begin{figure}[ht]
  \centering
  \includegraphics[width=0.9\linewidth]{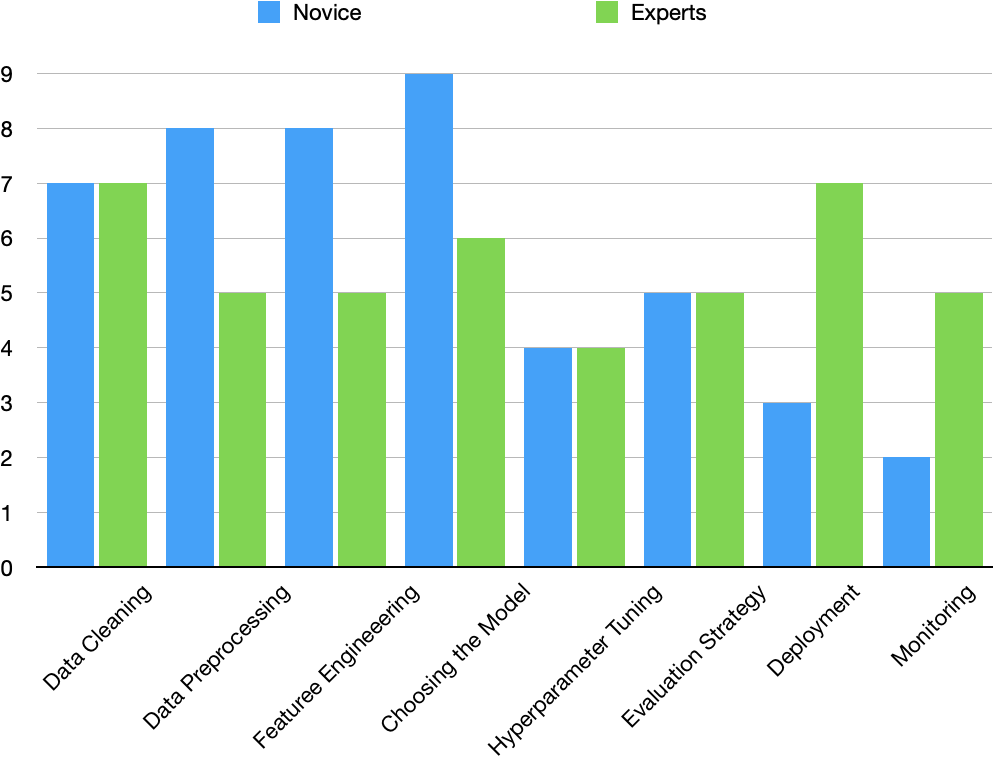}
  \caption{Tasks to be outsourced to ML experts.}
          \Description[Tasks to be outsourced to ML experts.]{
 A bar chart displaying the results of surveying the needs of GUI ML tools of novice and expert users. The data cleaning got Seven votes from novices and seven from expert users. Both data preprocessing and feature engineering got eight votes from novices and five from expert users. Choosing the model got nine votes from novices and six from expert users. Hyperparameter tuning got four votes from novices and four votes from expert users. The evaluation strategy got five votes from novices and five from expert users. Deployment got three votes from novices and seven votes from expert users. Monitoring got two votes from novices and five votes from expert users.}

  \label{Fig9}
\end{figure}

Furthermore, as shown in Fig.\ref{Fig9}, most novice users believe that data cleaning and preprocessing, model selection, and feature engineering Tasks should be outsourced to ML experts rather than working on them themselves. Expert users agreed with novice users when it came to data cleansing and model selection, and they also thought that model deployment should be outsourced. Only seven out of twenty-seven respondents provided reasons for outsourcing the chosen Tasks; time-consuming, lack of available tools, diversity of options, and focusing on more important Tasks are the reported reasons for outsourcing the chosen Tasks.
Regarding the characteristics that lead to usable tools, $28\%$ of novice users do not find that GUI ML tools provide necessary features, and $30\%$ of expert users agreed with them (where $75\%$ of them strongly disagree). None of the experts with more than one year of practice agreed that GUI ML tools are sufficient. Besides that, two of them cited a lack of data cleaning and preprocessing-related features. The survey results revealed that only $11\%$ of respondents disagree that ML tools must propose the most typical action. Furthermore, when the respondents were asked about what they liked most about the ML tool they used, eight of them praised the tool for its functionality aspects, seven praised the tool for its overall ease of use, and the others commended the availability of the tool as a free resource. However, the majority of respondents would prefer to modify the functionality of the ML tool design they've used. Others proposed improving the tool's assistance and documentation, as well as improving its user interface with additional usability features, including personalizing it based on the project, guiding abilities, and trust in the tool.


\subsection{GUI Machine Learning Tool Heuristics}
\label{Heuristics}
At this phase, we adapted Jakob Nielsen’s 10 Heuristics from 1994~\cite{Nielsen1990}, shown in Table~\ref{tab:heuristics-11}, to capture the most important issues when developing GUI ML tools for novice users based on our empirical evaluation.

{The heuristics proposed for GUI ML tools are shown in Table~\ref{tab:heuristics-22}. Heuristics 1, 3, 5, 6, and 8 were taken from Nielsen's set without modification as they were well-suited for GUI ML tools. The remaining heuristics (2, 4, 7, 9, and 10) were updated based on empirical evaluations. Heuristics 2 was updated by adding the word novice as during user testing and heuristic evaluation, as some of the issues uncovered could be familiar to an expert user but not a novice user. Heuristics 7 was also updated due to issues uncovered during user testing and heuristic evaluation by explicitly stating common shortcuts as users expected to use the universal shortcuts and were confused when they did not work (e.g., ctr(or cmd) z). Heuristics 4, 9, and 10 were updated based on user testing, survey results, and authors' experience during heuristic evaluation. The Cognitive Walkthrough and SUS results inspired the update of heuristics (4 and 9) and (4 and 10), respectively. As for Heuristic 4, the GUI tools evaluated used jargon associated with statistics or data mining in its early stages that is not understandable by novice users and does not even align with recent tutorials on ML problems. For Heuristics 9, the update was made because error messages and suggestions are more constructive when based on data, not just general errors (e.g., Illegal operation). Finally, for the last updated heuristics, Heuristics 10, observations during user testing show that users need to search YouTube to perform tasks repeatedly, and long hints provided by Weka made the participants impatient and jump from one label to another. Also, in the survey, an average of three processes need to be outsourced to experts to save time as they require background knowledge. 
}\par
{
As for the proposed heuristics in Table~\ref{tab:heuristics-22}, Heuristic 11 (\textbf{Guidance}) was suggested due to the need to follow a tutorial to build the model resulting from heuristic evaluation, cognitive walkthrough, emphasized by novice users in the survey, informal interviews after user testing, and SUS open-ended questions. \textbf{Trustworthiness} (Heuristic 12) is one of the needs mentioned by more than one respondent during the survey. In addition, trust and explainability are rising topics, and regulations are being implemented to enforce them~\cite{GDPR}. The necessity to balance between guidance and a simple aesthetic design needed by novices and the powerful endless functionalities that the user may need as the user grows with the system is the reason behind proposing the \textbf{adaptation to the growth} heuristic (H13), as was mentioned in the survey. Finally, the \textbf{context relevance} heuristic (H14) is treated first as an interpretation for other heuristics. However, it was later emphasized as an independent heuristic during informal interviews and user testing analysis conducted on the prototype in the fourth phase.}
\begin{table*}
\caption{{GUI ML Tools Heuristics.In \textbf{bold} proposed heuristics and updates to Nielsen's Heuristics}}
\label{tab:heuristics-22}
\resizebox{\textwidth}{!}{%
\begin{tabular} {p{0.3\textwidth}p{0.8\textwidth}} 
\toprule
GUI ML tools Heuristics                                                                                                                                                                                                                                                                                                                                           & Description                                                                                                                                                                                                                                                                                                                                                                             \\ \midrule
\textbf{1.Visibility of system status}                                                                                                                                            & The system should always keep users informed about what is going on through appropriate feedback within a reasonable time.                                                                                                                                                   \\\\
\textbf{2.Match between system and the real world}                    & The system should speak the users’ language, with words, phrases, and concepts familiar to the \textbf{novice} user rather than system-oriented terms. Follow real-world conventions, making information appear in a natural and logical order.
      \\ \\   
\textbf{3.User control and freedom}                                                                            & Users often choose system functions by mistake and will need a clearly marked "emergency exit" to leave the unwanted state without having to go through an extended dialogue. Support undo and redo.                                                                                                                \\\\

\textbf{4.Consistency and standards}                                                                                                    & Consistency and standards Users should not have to wonder whether different words, situations, or actions mean the same thing.\textbf{ Follow platform conventions and the current state of machine learning terms and jargon}.              \\\\

\textbf{5.Error prevention}   & Even better than good error messages is a careful design that prevents a problem from occurring in the first place. Either eliminate error-prone conditions or check for them and present users with a confirmation option before committing to the action.                                                            \\\\
\textbf{6.Recognition rather than recall}  & Minimize the user’s memory load by making objects, actions, and options visible. The user should not have to remember information from one part of the dialogue to another. Instructions for the use of the system should be visible or easily retrievable whenever appropriate. 
                                 \\\\
\textbf{7.Flexibility and efficiency of use}              & Accelerators – unseen by the novice user – may often speed up the interaction for the expert user such that the system can cater to both inexperienced and experienced users. Allow users to tailor frequent actions \textbf{and use the common shortcuts}.                     \\\\
\textbf{8.Aesthetic and minimalist design}                                                 & Dialogues should not contain information that is irrelevant or rarely needed. Every extra unit of information in a dialogue competes with the relevant units of information and diminishes their relative visibility.                                                                                     \\\\
\textbf{9.Help users recognize, diagnose, and recover from errors}                                                                                                    & Error messages should be expressed in plain language (no codes), precisely indicate the problem, and constructively suggest a solution,  \textbf{with respect to the user data.}                                                                                                   \\\\
\textbf{10.Help and documentation}         & Even though it is better if the system can be used without documentation,  \textbf{it may be necessary to provide tutorials and a real-time chatbox with}  \textbf{experts and} \textbf{concise hints}. Any such information should be easy to search,  focused on the user's task, list concrete steps to be carried out, and not be too large.  \\\\
\textbf{11.Guidance}                                                                                                      &\textbf{ The system should guide the novice user to the next appropriate step based on the problem and the data by suggesting a sequence of actions and providing recommendations.}                                                                                                     \\\\
\textbf{12.Trustworthiness}                                                                                                     & \textbf{The system should demonstrate trustworthiness by protecting user data and being truthful with the user. It should demonstrate to the user how the predictions were made.}                                                                                                  \\\\
\textbf{13.Adaptation to growth}                                                                                                     & \textbf{The system should adapt to novice user growth by reducing guidance, restrictions, and tips and allowing more customization and extendibility.}                                                                                                  \\\\
\textbf{14.Context relevance}                                                                                                    & \textbf{Display information relevant to the user’s current dataset and problem.}                                                                                                   \\
                                                                                                                                                                                                                                                                                                                                                            & \\ \bottomrule
\end{tabular}%
}
\end{table*}


\section{Phase 2: Develop a prototype that meets the new proposed heuristics}
\label{Ph2}
By using Weka as a baseline with extra usable features, a prototype\footnote{{In the supporting material}} for more usable GUI ML tools was produced in this phase that satisfies the new proposed heuristics resulting from the previous phase. The four steps of the usability engineering lifecycle presented in~\cite{bahr2017prototyping} were followed to produce our prototype sequentially, including analysis of requirements, prototype design, prototype creation, user-test/expert evaluation. These steps were repeated in a new iteration as needed. In the first step, the requirements were identified (e.g., such as the need to load the dataset, perform exploratory data analysis, train the mode, and evaluate the model) { with a focus on the $13$ heuristics}~\footnote{{ The 14th heuristic mentioned in the table was added during the fourth step of the process}} as stated in Table~\ref{tab:heuristics-22}. 
Also, by considering Weka's functionalities refined by going through multiple tutorials for novice users to focus only on the basic functionalities.\par
\begin{figure*}[ht]
  \centering
  \includegraphics[width=0.7\textwidth]{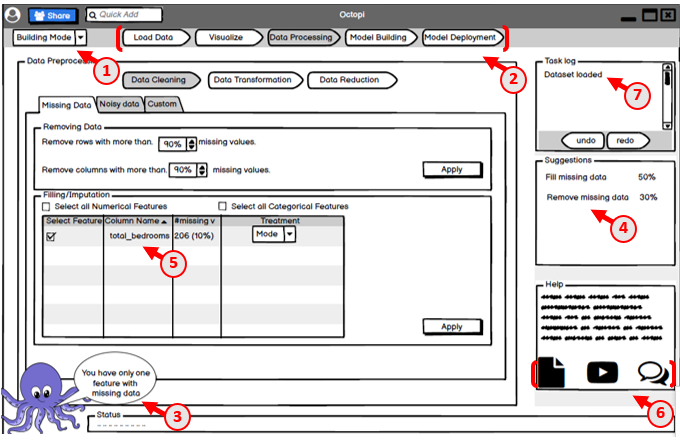}
  \caption{Data preprocessing window of the prototype shows some of the heuristic's features: (1) build mode or deployment mode drop-down list, (2,3,4) guidance and suggestions, (5) displaying information relevant to the user's current dataset, (6) different types of help (tutorials, a real-time chatbot with experts, and documentation), (7) the tasks performed by the user.}
  \label{Fig18}
\end{figure*}

Following that, in the second step of the usability engineering lifecycle, a prototype was designed to address the identified requirements. As illustrated in Fig.~\ref{Fig18}, the guidance heuristic was implemented in general by displaying the formal sequence for using any ML tool in the form of sequential taps on the top of the interface, additionally by presenting a suggestions box that presents the main tasks regarding each interface the user can perform. The guidance heuristic was also supported by supplying the interfaces with an agent, the octopus, who guides the user by giving comments regarding the essential tasks that can be done through the interface. We chose the octopus as a guidance agent in our prototype since it has the largest brain of any invertebrate. Moreover, the help heuristic was applied in the prototype by providing tutorials and a real-time chatbox with experts in addition to the documentation. Another usable feature is the task log box that presents the tasks performed by the user with undo and redo buttons to support the user's control and freedom heuristic.
In the model-building window, we added an explainability tab to emphasize the system's trustworthiness heuristic to the user by showing how the prediction was made. We also reviewed multiple papers, websites, and blogs to ensure that the terminologies conform to the ML field, supporting the Consistency and standards heuristic. {The adaptation to  growth heuristic cannot be tested through the prototype as it requires a longitude study on a fully functioning software; however, it was incorporated by adding a custom tab for each module in which users can add code directly. Also, the guidance provided by the octopus will decrease in terms of the frequency of the learned items.}\par
After that, in the third step, the prototype was created using one of the most widely-used prototyping tools in academia, Balsamiq~\cite{Balsamiq75,kim2021improving}. The prototype has two modes: (1) Building mode that caters to Loading and preprocessing the Data and Building and Exporting the Model (Tasks 1-6.a). (2) Deployment mode in which, independently from the Building mode, the user can load a built model and make predictions (Tasks 6.b-6.d). 
Once the prototype was created as an interactive UI that responds to the user actions, it was tested in the fourth step through expert evaluations initially, where domain experts provided feedback about the interface implementation with Nielsen's heuristics, shown in Table~\ref{tab:heuristics-11}. Then, it was used to evaluate the new proposed heuristics in-depth with real users by conducting a user testing evaluation as discussed in the next Section~\ref{Ph3}.

\section{Phase 3: Validation of developed heuristics}
\label{Ph3}
\subsection{Experimental Design}
\label{Ph3-Experimental}
After developing a prototype for a GUI ML tool (by considering Weka as a baseline) that meets the new proposed heuristics, we conducted a user testing evaluation of that prototype in order to validate the proposed heuristics. The evaluation procedure was identical to the user testing evaluation for Weka described in section~\ref{Ph1-Experimental} with another group of thirteen participants recruited through personnel connections.
\subsection{Results and Analysis}
\label{Ph3-Results}
\subsubsection{User testing evaluation of prototype.}
Similar to section~\ref{Ph1-Experimental}, our goal here was to measure the proficiency and error rate when performing the same set of $12$ tasks on the prototype developed that follows the proposed heuristics for evaluating GUI ML tools. Thirteen novice participants completed the experiment. The results of this experiment indicate a highly usable system with $100\%$ success rate across the $12$ Tasks, an average completion time for all the Tasks of $11$ minutes, and an average completion time of $55$ seconds per task. Fig.\ref{Average-Task-completion-duration-p2} (in the appendix) illustrates the breakdown of the different tasks. The minimum average was about $16$ seconds for Task 5 (model evaluation), while the maximum average completion duration for a task was one minute and $36$ seconds for Task 3c (splitting the data to training and testing) followed by a one second difference in Task 6b (loading the saved model). \par
Task 6b had the highest error, followed by Task 3c,  illustrated in Fig.\ref{Error-rate-p2} (in the appendix). The cause of the error in Task 3c was that novice users found splitting the data was better to be grouped under data preprocessing rather than model building. As in Task 6b, the errors were caused by the confusion between the \emph{Model Deployment} tab and \emph{Deployment Mode} option in the drop-down list. An error by confusion class is the source of errors for the other Tasks also due to similar terminology. It was remarkable that $4$ of the $12$ classes were error-free or had an average error less than $0.1$.

\subsubsection{SUS analysis for the proposed prototype.}
\begin{figure}[ht]
  \centering
  \includegraphics[width=0.9\linewidth]{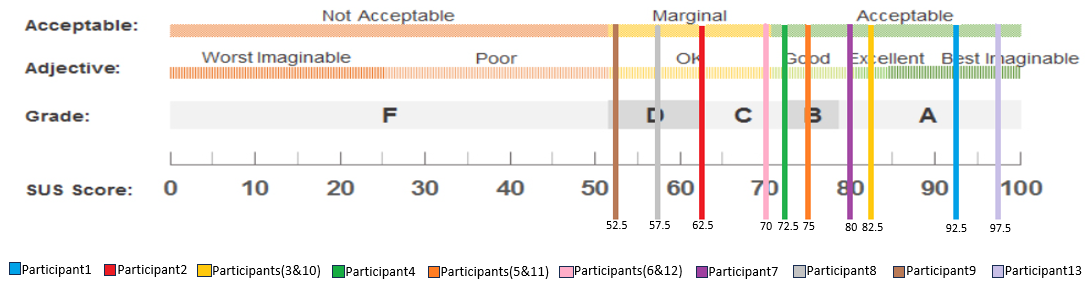}
  \caption{SUS evaluation of the prototype.}
\Description[SUS evaluation of the prototype.]{
 A diagram showing the SUS scores of thirteen participants with the prototype. The SUS scores of participants 1, 2, 3, 4,5, 6, 7, 8, 9, 10, 11, 12, and 13 are 92.5, 62.5, 82.5, 72.5, 75,70, 80, 57.5, 52.5, 82.5, 75,70, and 97.5, respectively.}

  \label{sus-p2}
\end{figure}

Fig.\ref{sus-p2} demonstrates the proposed prototype satisfaction measures regarding the SUS score interpretation of thirteen participants. It was measured and analyzed in a similar way to Weka. According to the acceptance category, Eight participants (1, 3, 4, 5, 7, 10, 11, and 13) accepted the prototype with a high level of usability, ranging from 'good' to 'best imaginable' with 'A' and 'B' grades, indicating that the prototype has a high performance. However, the remaining participants (2, 6, 8, 9, and 12) reported a marginal satisfaction level for the prototype with 'OK' usability. Participants 2, 6, and 12 reported a 'C' grade, indicating an average level of satisfaction, while participants 8 and 9 reported a 'D' grade, which means their satisfaction level was less than average. On average, the prototype achieved a SUS score of 74.62, which is considered acceptable according to Table~\ref{tab:sus-range}.

When it comes to the open-ended questions, participants did agree that the octopus was helpful, and the sequence of actions was clear. One participant elaborated further on that by saying, "The most thing I like about the tool is that I became more familiar with what I am doing and what my next steps are. I believe this was accomplished through the horizontal bar that identifies where I am and what I am doing, the task log on the right, which summarises what I have done, and the little octopus that guides me along". As for what to improve, two participants did not offer suggestions. One participant mentioned similar terminology that caused some confusion. Another participant suggested changing the position of the octopus and finally adding some colors and design elements.

\section{Phase 4: Revising the Heuristics}
\label{Ph4}
After generating the heuristics in Phase 1, heuristics 1-13 in Table~\ref{tab:heuristics-22}, 
then validating them by conducting user testing of the developed prototype, we revised them to ensure their effectiveness considering the error types that the users made from Phase 3. As the error source of the errors mentioned in section~\ref{Ph3-Results}, was not due to limitations of the proposed heuristics, there is no need to modify the adapted heuristics. Diving deeper into the cause of the steep drop in error rate, and taking as an example Task 3a: Fill in missing data, one of the reasons for this result is that in the prototype, we provided the user with only the feature that contained missing data and needs to be filled. A similar approach was followed in other tasks and for other types of information, and this approach was highlighted by the participants in the informal interviews. Thus, we added to our proposed heuristics the fourteenth heuristic, which is Context Relevance, with the description of "Display information relevant to the user's current dataset and problem". As a result, the final set of our proposed heuristics consists of the $14$ heuristics presented in Table~\ref{tab:heuristics-22}. 
\section{Discussion}
\label{Dis}
The GUI ML tools team has done a commendable job by providing a range of data science-related functionalities in a GUI format that users can access without coding. However, based on several usability evaluations conducted on two of the most commonly used GUI ML tools (WEKA and KNIME) to answer (RQ1), it was found that the current GUI ML tools have significant and catastrophic usability issues that need to be addressed urgently before the product can be released. 
Some frequent problems identified include the need for better feedback, adequate help and documentation, and too much unnecessary information and options that may need to be clarified for users. Additionally, the tools require clear labeling and buttons to minimize confusion. During user testing, all errors were categorized as errors by confusion, with the top source of confusion being the lack of clear labels or buttons and missing paths. 
Regarding the satisfaction usability attribute, WEKA, one of the most popular GUI ML tools, received an average unacceptable satisfaction score of 49. As a result, the usability of those tools needs to be improved.
Additionally, it was found that users faced difficulty understanding the interface, requiring more explicit guidance, learnability, trust, functionality, and aesthetic aspects. This led to the need for a more comprehensive set of heuristics. In 1990, Nielsen and Molich developed usability guidelines, but since then, user interfaces have evolved, and researchers have recognized the need to adapt Nielsen's broad set of heuristics to specific interfaces ~\cite{langevin2021heuristic}. In this research, to address RQ2, we have proposed new heuristics and updates to Nielsen's heuristics, which can be applied to GUI ML tool interfaces to provide a good user experience using a four-phased approach.
In phase one, we generated an initial set of heuristics based on Nielsen's heuristics, using the results of a conducted cognitive walk-through and Nielsen's heuristic evaluation of two popular GUI ML tools (Weka and KNIME), surveying the needs of GUI ML tools users, and user testing evaluation of Weka. This resulted in three new heuristics and five updated heuristics. It is worth noting that the proposed heuristics, although for us originating from the empirical evaluation, were also mentioned when adapting Nielsen's heuristics to other domains or proposing Usability Guidelines. Guidance and trustworthiness (Heuristics 11 and 12) were proposed in the heuristics for conversational agents~\cite{langevin2021heuristic}. In the conversational agents' context, guidance was a result of two heuristics, \emph{Recognition rather than recall} and \emph{Help and Documentation}, in which users should be provided with feedback and guidance throughout the conversation to understand the status of the system better. For our context, it is more about guiding the user to what he should do next, either in terms of the upcoming tasks or how to perform the current task~\cite{langevin2021heuristic}. It also presents an alternative to scaffolding, heavily used in current iML tools, and follows a suggestion made in~\cite{Mishra2021} that instead of scaffolding a recommender system that specifically focuses on suggesting helpful next steps based on the user's prior steps in the transfer learning workflow pipeline may assist non-expert users in their progress without the limitations imposed by scaffolding.\par
As for trustworthiness, the conversational agent should be truthful in its interactions to encourage trustworthiness. However, in our context, the interaction is not the dialog~\cite{langevin2021heuristic}. It is related to how the data is stored and how the predictions are made. As for Adaptation to growth (Heuristic 13), It was proposed in the context of providing Guidelines to Human-AI Interaction~\cite{amershi2019guidelines}. Adaptation to growth has a similar meaning to the "Learn from user behavior" guideline, in which in the context of Human-AI Interaction, AI should personalize the experience by learning from the human actions over time~\cite{amershi2019guidelines}, whilst in the context of GUI ML tools we personalize the experience by adapting to the learning curve of the user and reducing the intensity of the tips and recommendations as to the learning increases. In our context, the system should allow for customization and extendability with Python code, for example, as the user gains experience.

\begin{figure}[ht]
  \centering
  \includegraphics[width=0.9\linewidth]{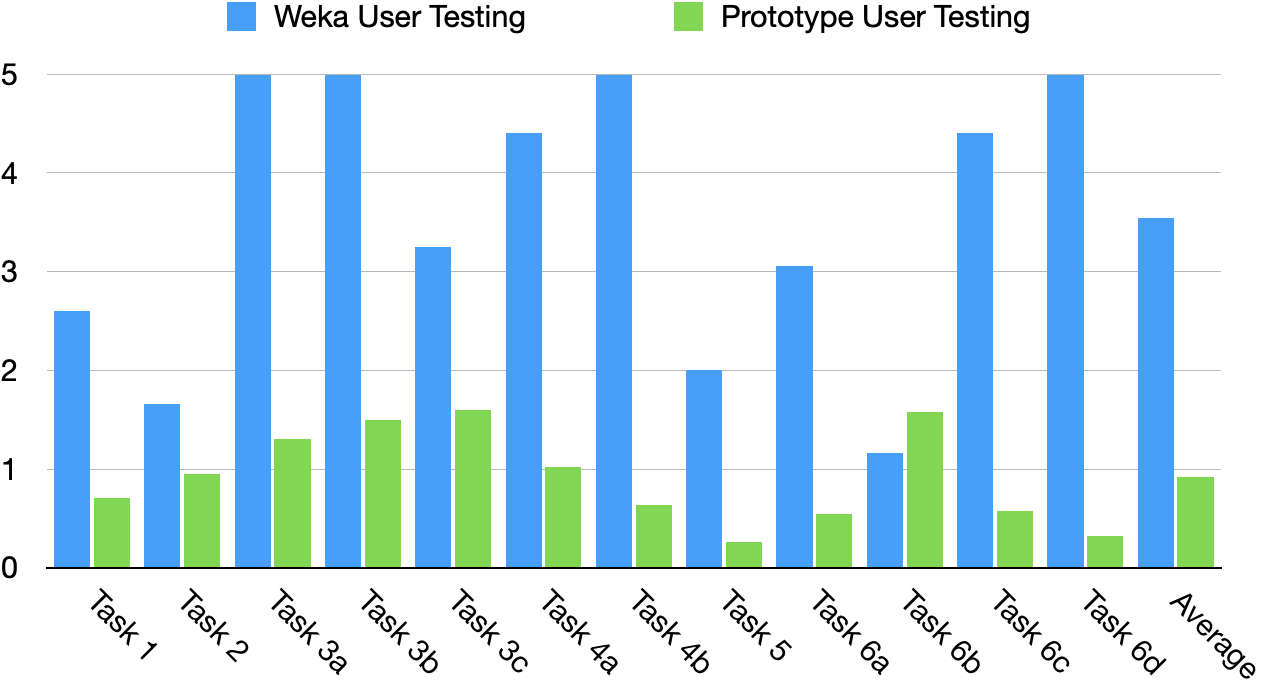}
  \caption{Tasks Completion Comparison \label{Tasks-Completion-c}.}
            \Description[Tasks Completion Comparison.]{
 A bar chart displaying the results of comparing Task completion duration of user testing of WEKA and the prototype. For WEKA, Tasks 1, 2, 3a, 3b, 3c, 4a, 4b, 5, 6a, 6b, 6c, and 6d have a Task completion duration of 2.60, 1.66, 5.00, 5.00, 3.25, 4.40, 5.00, 2.00, 3.06, 1.16, 4.40, and 5.00, respectively, and the overall average is 3.54. For the prototype, Tasks 1, 2, 3a, 3b, 3c, 4a, 4b, 5, 6a, 6b, 6c, and 6d have a Task completion duration of 0.71, 0.95, 1.31, 1.50, 1.60, 1.02, 0.64, 0.27, 0.54, 1.54, 1.58, 0.57, and 0.32, respectively, the overall average is 0.92.}

\end{figure}

\begin{figure}[ht]
  \centering
  \includegraphics[width=0.9\linewidth]{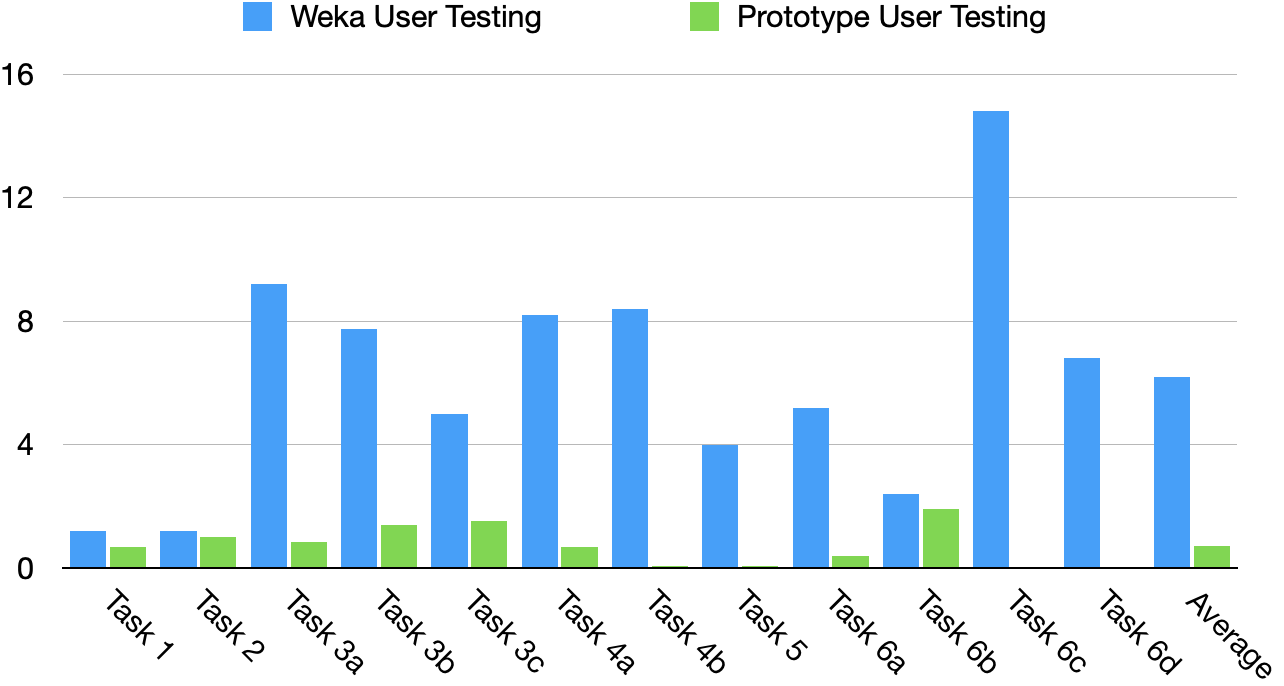}
  \caption{Error rate Comparison \label{Error-rate-c}.}
          \Description[Error rate Comparison.]{
 A bar chart displaying the results of comparing Task completion duration of user testing of WEKA and the prototype. For WEKA, Tasks 1, 2, 3a, 3b, 3c, 4a, 4b, 5, 6a, 6b, 6c, and 6d have an error rate of 1.20, 1.20, 9.20, 7.75, 5.00, 8.20, 8.40, 4.00, 5.20, 2.40, 14.80, and 6.80, respectively, and the overall average is 6.18. For the prototype, Tasks 1, 2, 3a, 3b, 3c, 4a, 4b, 5, 6a, 6b, 6c, and 6d have an error rate of 0.69, 1.00, 0.85, 1.38, 1.54, 0.69, 0.08, 0.08, 0.38, 0.38, 1.92, and 0.00, respectively, the overall average is 0.72.}

\end{figure}

Then, in phase two, we developed a prototype based on the newly proposed heuristics. Then, we validated the proposed heuristics in phase 3, using user testing on the developed prototype. Phase 3 showed significant improvement in task completion and reduction of error rates as comparing the results of user testing of Weka~\ref{Ph1-results} and user testing of the prototype, the superior performance of the prototype built based on the heuristics over Weka is obvious, with a completion rate 3.9x faster in favor of the prototype, illustrated in Figure~\ref{Tasks-Completion-c}. In addition, as illustrated in Figure~\ref{Error-rate-c}, the error rate was ninth of the errors captured when user testing Weka. The errors are narrowed down to using similar terminology such as 'deployment mode' and 'model deployment', which was a mistake from our side and violated Nielsen's Forth Heuristic. Thus, we replaced the Model Deployment button label with Export the Model. It was interesting too that although splitting the data is usually grouped with modeled building, in Weka, for example, for efficiency of use, novice users did suggest having it with data preprocessing favoring consistency and standard.\par

Phase 3 also showed an increase in user satisfaction as the SUS survey went from a score of $49$ for Weka to a score of $74.62$. Moreover, conversations in the informal interviews and SUS open-ended questions from "I feel I'm lost", and "The labels are confusing, I cannot link them with the tutorial", to the need to add delighters such as colors, motion to the octopus, all of the aforementioned adds to the validity of the generated heuristics. \par

Finally, in phase 4, we revised the new heuristics by proposing an additional heuristic, raising the total number of the GUI ML tools heuristics to 14. Context Relevance was also proposed in the context of providing Guidelines to Human-AI Interaction and maps to Show contextually relevant information in the study~\cite{amershi2019guidelines}, with the difference that in Human-AI Interaction, the context is the user's inferred goals and attention during the interaction, while in our GUI ML Tools Heuristics, the context is the current dataset and problem. This becomes increasingly important when building Fair ML models as emphasized by practitioners~\cite{10.1145/3581641.3584064}. To summarize, after conducting a usability evaluation on the prototype of the GUI ML tool that followed the newly proposed set of heuristics, it was found that fewer usability issues were present in the prototype. This indicates that the proposed set is more appropriate for assessing the usability of GUI ML tools than Nielsen's set (RQ2).

\par


\subsection{Threats to Validity}
\textbf{External validity} is threatened by two key factors. First, just two GUI ML tools (Weka and KNIME) were investigated and used in the experiment to identify usability issues with GUI ML tools. To mitigate this threat, Weka and KNIME were tested using the three most common usability evaluation methods: cognitive walkthrough and heuristic evaluation for both tools, as well as user testing, including SUS for Weka. Second, the user testing evaluation for Weka had a limited sample number of participants. As a result, these users may not be a representative sample of all novices. Furthermore, while 27 responses to the survey conducted to collect user needs for GUI ML tools is a reasonable number, additional responses are required to obtain more needs and perceptions.
\textbf{Construct validity} is threatened by the fact that the developed prototype didn't meet the adaptation to the growth heuristic. So, the conducted user testing cannot be used to validate this heuristic because it tends to be presented to the novice user after a long period of usage; therefore, it is impossible to validate it from a single user testing session. 
In terms of the \textbf{Conclusion validity}, the results of this study are adequate to conclude. The intended users and tasks were carefully gathered through user and task analysis. Furthermore, the usability measures were chosen concerning the novice users, and the GUI ML tools were selected based on their popularity and significant features. Finally, all the data were made available for replication, and the methodology of this study is detailed in Section 3.

\section{Conclusions}
\label{conc}
With the rise of ML techniques and them being integrated with many disciplines where professionals are not familiar with the basic concepts of ML or computation techniques in general, the need for usable GUI ML tools arises. However, when addressing RQ1, critical usability issues were identified when evaluating popular GUI ML tools, including inadequate feedback, documentation, and excessive and confusing options. This resulted in an unsatisfactory satisfaction score, emphasizing the need to improve these tools' usability. Therefore, this work provided a set of 14 heuristics adapted and customized from Nielsen's heuristics. This set was developed through a series of empirical evaluations to generate the heuristics. The evaluations included Nielsen's heuristics evaluation, cognitive walkthrough, a survey, user testing, and SUS survey. Upon generating the heuristics, a prototype was developed conforming to those heuristics, and through user testing, those heuristics were validated and revised. Based on the results, it is evident that the prototype designed using the proposed heuristics performs better in terms of usability when compared to the GUI ML tool (WEKA). This indicates that the new heuristics are more effective in enhancing the usability of GUI ML tools and user experience by including new heuristics such as Guidance, Trustworthiness, Adaptation to growth, and Context relevance (RQ2). Furthermore. We expect this work will lay a solid foundation for designers seeking to develop GUI ML tools with a delightful user experience to add to the toolkit of professionals across disciplines.\par
The future direction of this work has three folds: a Machine Learning part concerning building the logic and models behind the suggestion system, the personalization, and the octopus helper. The basic tasks that cater to novice users across various problems should be systematically investigated in cross-disciplinary work between requirement engineering and Machine Learning. Lastly, in terms of usability, a longitude experiment should be conducted to measure the user experience in terms of trust and adaptation to growth, as they are hard to measure from a single experiment. 
\begin{acks}
The authors acknowledge the support of King Fahd University of Petroleum and Minerals in the development of this work and acknowledge the support provided by the Deanship of Research Oversight and Coordination at King Fahd University of Petroleum \& Minerals (KFUPM). The authors would like to thank all participants in the study.
\end{acks}

\bibliographystyle{ACM-Reference-Format}
\bibliography{main}

\appendix
\section*{Appendix}
\begin{table*}
\caption{Jakob Nielsen's severity rating.}
\label{tab: severity}

\begin{tabular}{@{}ll@{}}
\toprule
\textbf{Severity} & \textbf{Description}                                                                \\ \midrule
0                 & I don't agree that this is a usability problem at all                               \\
1                 & Cosmetic problem only: need not be fixed unless extra time is available on the project. \\
2                 & Minor usability problem: fixing this should be given low priority.                  \\
3                 & Major usability problem: important to fix, so should be given high priority.        \\
4                 & Usability catastrophe: It is imperative to fix this before the product can be released.       \\ \bottomrule
\end{tabular}%

\end{table*}
\begin{table*}
\caption{Tasks for user testing evaluation}
\label{tab: user-testing-q}
\resizebox{0.8\textwidth}{!}{%
\begin{tabular}{@{}ll@{}}
\toprule
Task id & Tasks for user testing evaluation                                                                        \\ \midrule
1 & 1. Load the dataset from a file named 'Housing.csv'                                                            \\
2 & 2. Visualize the data by presenting the correlations between the number of rooms and the latitude features.\\
3a & 3. Fill in missing data by taking the mode.                                                                     \\
3b & 4. Perform one hot-encoding on categorical data (using NominalToBinary transformation).                         \\
3c & 5. Prepare the data for training and evaluation by splitting the data (70\% training and 30\%  testing).              \\
4a & 6. Build the model using a Random Forest Regressor; the target value is the median house value feature.   \\                 

4b & \begin{tabular}{@{}l@{}} 7.Perform Hyperparameter tuning by changing the max depth 50 and estimators (trees) to 50. 
\end{tabular} \\
5 & 8. Find the Mean Absolute Error of the prediction.                                                              \\
6a & 9. Save the model with the name 'RFhousing.'                                                                    \\
6b & 10.Load the saved model.                                                                                       \\
6c & 11. Perform the prediction on the 'dep.csv' file.                                                                   \\
6d & 12. Find the results of the prediction.                                                                         \\ \bottomrule
\end{tabular}%
}
\end{table*}
\begin{table*}
\caption{Results of Heuristic Evaluation of Weka and KNIME}
\label{tab:HE-results}

\resizebox{\textwidth}{!}{%
\begin{tabular}{@{}lllll@{}}
\toprule
\textbf{H\#}         & \textbf{Issue}                                                                                                                                                                                                                                                                                                  & \textbf{Weka}                                                    & \textbf{KNIME}                                                                & \textbf{Severity} \\ \midrule
\multirow{7}{*}{1}  & \begin{tabular}[c]{@{}l@{}}1.1 When the file(dataset) loaded /filter was applied for (One-hot encoding/ missing data)\\  no related feedback is shown immediately (in the status)\end{tabular}                                                                                                                  & \begin{tabular}[c]{@{}l@{}}1\\ 3.a,3. b\end{tabular}             &                                                                               & 2                 \\
                    & 1.2 No feedback was provided when the model saved/loaded (in the status)                                                                                                                                                                                                                                        & 6.a,6. b                                                         &                                                                               & 2                 \\
                    & 1.3 There is no clear feedback on what a user should do after dragging a node into the space                                                                                                                                                                                                                    &                                                                  & Across Tasks                                                                  & 3                 \\
                    & 1.4 No clear feedback after loading the data, or building a chart on how to view it or visualize it                                                                                                                                                                                                             &                                                                  & 1, 2                                                                          & 4                 \\
                    & \begin{tabular}[c]{@{}l@{}}1.5 There is feedback that some operation was performed by changing the color to green,\\  but no feedback on exactly what happened which is not sufficient for many Tasks.\end{tabular}                                                                                             &                                                                  & Across Tasks                                                                  & 2                 \\
                    & 1.6 There is no clear way to know, after partitioning the data, which part of the output is which split(train/test)                                                                                                                                                                                             &                                                                  & 3.c                                                                           & 3                 \\
                    & \begin{tabular}[c]{@{}l@{}}1.7 When attempting changes, it tells the user that the node(s)\\  will be reset without telling me which nodes will get affected or how\end{tabular}                                                                                                                                &                                                                  & 3.a                                                                           & 3                 \\

2                   & 2.1 The view button can be confused with a search button in terms of metaphor                                                                                                                                                                                                                                   & 2                                                                &                                                                               & 2                 \\
3                   & 3.1 There is no undo after performing a transformation                                                                                                                                                                                                                                                          & 3.a,3.b                                                          & 3.a,3.b                                                                       & 3                 \\
\multirow{2}{*}{4}  & 4.1 The label of the button (Set...) is not expressive of its function                                                                                                                                                                                                                                          & 6.c                                                              &                                                                               & 2                 \\
                    & 4.2 The terminology used is not consistent with the terminology used in the ML.                                                                                                                                                                                                                                 & 3.a                                                              & \begin{tabular}[c]{@{}l@{}}6.a, 5, 4.a,3.c,\\ 3.b, 3.a 6.d, 6.c\end{tabular} & 3                 \\
                    \multirow{3}{*}{5}  & 5.1 No error prevention (warning) provided when choosing unsupported file type                                                                                                                                                                                                                                  & 1                                                                & -                                                                             & 3                 \\
                    & \begin{tabular}[c]{@{}l@{}}5.2 No error prevention (lists) provided to prevent the user from inputting wrong values in \\ (attributeIndices) text box for configuring the selected filter, in addition, it is not clear how\\  to apply the filter to the attributes other than the first and last\end{tabular} & 3.a                                                              & -                                                                             & 3                 \\
                    & 5.3 Allows the user to apply regression models when the target is binary                                                                                                                                                                                                                                        & 4.a                                                              & 4.a                                                                           & 3                 \\
                    & \begin{tabular}[c]{@{}l@{}}5.4 does not chose proper default target value to prevent\\  the evaluating on the wrong attributes, The columns chosen by default were not correct\end{tabular}                                                                                                                     &                                                                  & 5                                                                             & 4                 \\
                    & 5.5 No error prevention method for connecting unreasonable training and testing splits                                                                                                                                                                                                                          &                                                                  & 3.c                                                                           & 3                 \\
                    & 5.6  does not restrict the user from trying to draw the chart while there is no data                                                                                                                                                                                                                            &                                                                  & 2                                                                             & 1                 \\

6                   & 6.1 Too many steps to reach to perform a simple Task                                                                                                                                                                                                                                                            & 3.a,                                                             & \begin{tabular}[c]{@{}l@{}}2, 3.a, 3.b, 3.c,\\  6.b, 6.c,6.d\end{tabular}     & 3                 \\
\multirow{2}{*}{7}  & 7.1 The tool does not support common open shortcut (Ctrl+O) to show open file dialog box                                                                                                                                                                                                                        & 1                                                                & 1                                                                             & 1                 \\
                    & 7.2 The user can only perform the hyperparameter tuning manually                                                                                                                                                                                                                                                & 4.b                                                              & 4.b                                                                           & 2                 \\
\multirow{13}{*}{8} & 8.1 No need for Invert button (Boxes that are ticked become unticked and vice versa)                                                                                                                                                                                                                            & 2                                                                &                                                                               & 1                 \\
                    & 8.2 To configure the selected filter, the user must click on the filter name, which is not clear the user                                                                                                                                                                                                       & 3.a                                                              &                                                                               & 3                 \\
                    & 8.3 To configure the model the user must click on its name which is not clear for the user                                                                                                                                                                                                                      & 4.b                                                              &                                                                               & 3                 \\
                    & \begin{tabular}[c]{@{}l@{}}8.4 To save/ load the model the user must right click\\  on the white space under “Result list” which is not clear for the user\end{tabular}                                                                                                                                         & \begin{tabular}[c]{@{}l@{}}6.a\\ 6.c\end{tabular}                &                                                                               & 3                 \\
                    & \begin{tabular}[c]{@{}l@{}}8.5 The label of the button(close) in the test instances window (for choosing the test set)\\  is not expressive of its function, it should be replaced with 2 buttons ok and cancel\end{tabular}                                                                                    & 6.c                                                              &                                                                               & 2                 \\
                    & 8.6 too many options to load or save a file many that are not commonly used, or obsolete                                                                                                                                                                                                                        &                                                                  & \begin{tabular}[c]{@{}l@{}}1, 6.a,\\  6.b, 6.d\end{tabular}                   & 3                 \\
                    & 8.7 No simple dialog or path for visualizing the data                                                                                                                                                                                                                                                           &                                                                  & 2                                                                             & 2                 \\
                    & 8.8  The way of grouping the transformation or hyperparameters is not convenient.                                                                                                                                                                                                                               &                                                                  & 3.a, 4.b                                                                      & 2                 \\
                    & 8.9 The presence of 2 icons performing the same Task                                                                                                                                                                                                                                                            &                                                                  & 2                                                                             & 1                 \\
                    & 8.10 Shows all target column with disregards of its applicability to the model                                                                                                                                                                                                                                  &                                                                  & 4.a                                                                           & 3                 \\
                    & \begin{tabular}[c]{@{}l@{}}8.11 Multiple outputs for the trained model, which makes\\  it unclear to decide which output the user should connect with the predictor\end{tabular}                                                                                                                                &                                                                  & 4.a                                                                           & 4                 \\
                    & \begin{tabular}[c]{@{}l@{}}8.12 When opening the window for opening or  saving a file, it goes to the root path\\  of the device instead of the workspace or most recently opened\end{tabular}                                                                                                                  & \begin{tabular}[c]{@{}l@{}}1, 6.a,\\  6.b, 6.c, 6.d\end{tabular} & \begin{tabular}[c]{@{}l@{}}1, 6.a, 6.b, \\ 6.c, 6.d\end{tabular}              & 3                 \\
                    & 8.13 Reading models and reading data are in the same list                                                                                                                                                                                                                                                       &                                                                  & 1, 6.b                                                                        & 3                 \\
\multirow{9}{*}{9}  & \begin{tabular}[c]{@{}l@{}}9.1 No error message on what went wrong or what to do\\  when the user inputting an illegal hyperparameter value\end{tabular}                                                                                                                                                        & 4.b                                                              & 4.b                                                                           & 3                 \\
                    & 9.2 Un-clear error message when choose wrong destination(folder) to save the model                                                                                                                                                                                                                              & 6.a                                                              &                                                                               & 3                 \\
                    & 9.3 Un-clear error message when choosing wrong model file                                                                                                                                                                                                                                                       & 6.b                                                              &                                                                               & 3                 \\
                    & 9.4 Un-clear error message after choosing a test set to perform the prediction                                                                                                                                                                                                                                  & 6.c                                                              &                                                                               & 3                 \\
                    & 9.5 does not show adequate error message when selecting an unsupported file                                                                                                                                                                                                                                     & 1                                                                & 1                                                                             & 4                 \\
                    & 9.6 error messages don't show how to solve the problem,                                                                                                                                                                                                                                                          &                                                                  & 2                                                                             & 4                 \\
                    & \begin{tabular}[c]{@{}l@{}}9.7 no warning message if connecting unreasonable training\\  and testing splits to training and prediction\end{tabular}                                                                                                                                                             &                                                                  & 3.c                                                                           & 2                 \\
                    & 9.8 No warning message when choosing a nominal value for regression                                                                                                                                                                                                                                             &                                                                  & 4.a                                                                           & 2                 \\
                    & 9.9 Unclear error message when the scorer is unable to score                                                                                                                                                                                                                                                    &                                                                  & 5                                                                             & 3                 \\
10                  & 10.1 No offline help, No proper help regarding any Task                                                                                                                                                                                                                                                         & all Tasks                                                        & all Tasks                                                                     & 3                 \\ \bottomrule
\end{tabular}%
}
\end{table*}

\begin{table*}
\caption{Cognitive walkthrough results.Task X Action Y will be referred to as Task X A(Y).}
\label{tab:CWR}
\resizebox{0.83\textwidth}{!}{%
\begin{tabular} {p{0.5\textwidth}p{0.1\textwidth}p{0.2\textwidth}p{0.3\textwidth}} 
\toprule
 Issue         & Weka & KNIME                                                                                                                                                        & Suggestion \\ \midrule
 
  \multicolumn{4}{l}{Issues related to question 2: Will the user notice that the correct action is available?} \\
 
 \midrule
\textbf{Issue 1: Unclear start of the ML development workflow} When opening the tools, there is no clear guidance on from where to start & Task 1 A(1) & Task 1 A(1) & The explorer window should open immediately, with an option to navigate between Weka's other functionalities later, an example ML workflow should be present once the user open the tool.                                                                                                                                                \\
\textbf{Issue 2: The user must click on the filter name to configure it} This issue is present in the preprocessing tab, when the user wants to perform some preprocessing to the data  & Task 3. A(3) & - & Open the configuration pop-up window immediately after selecting the filter by the user. Also, add a "Configure" button beside the "Apply" button to have the ability to reconfigure.                                                                                                                                                \\
\textbf{Issue 3: The user must right-click/secondary click on the white space to save/load a model} This issue is presented when the user attempts to save the model after training or load it to perform the prediction  & Task 6.a/6.b A(1) & - & Add a clear drop-down list for more options such as (Save/Load) the model                                                                                                                                                \\
\textbf{Issue 4: Tab label overload} The user must click on the classify tab to prepare the data for training and evaluation, perform regression, and evaluate the model. The naming of (Classify) is very specific, making the user feel lost.& Task 3.c A(1), Task 4.a A(1), Task 5 A(1) & - & Change the name of the tab (Classify) to something more general such as "Model building", as this tab holds more functionalities than just applying classification.                                                                                                                                                \\
\textbf{Issue 5: The needed Action is under multiple categories or steps}  This issue is present in KNIME, where the needed action is under multiple steps and is not obvious to the user. This issue is also present in Weka when it comes to exporting the prediction results.  & Task 6.d A(1) & Task 1, A(1,3), Task 2 A(1,4), Tasks 3.a,3.b(2),  Task 4.a A(2), Task 6.a A(1), Task 6.b A(1,3), Task 6.c A(3) and Task 6.d A(1) & Important actions must be performed in the minimal number of steps.                                                                                                                                                \\
\midrule
   \multicolumn{4}{l}{Issues related to question Q3:  Will the user associate the correct action with the effect that the user is trying to achieve?} \\
 
 \midrule

\textbf{Issue 1: The user may select unsupported dataset file type} Weka allows the user to select unsupported file type or provide any feedback that indicates so. This may mislead the user that they can load the unsupported file type safely. & Task 1 A(3) & - & Disable selecting unsupported file formats. In addition, show a pop-up warning informative message to notify the user that the selected type is unsupported.                                                                                                                                                \\
\textbf{Issue 2: Unclear label for menus}
In WEKA, the user may not associate the word filter with preprocessing techniques to reach the panel. Also, exporting or saving the prediction results is via a 'chosen' button with a label next to it of 'output prediction' and a placeholder that contains 'null', which is unclear. Also, in KNIME, loading or saving a file or models under IO, preprocessing techniques under manipulation, training, and testing under analytics & Tasks 3. a, 3. b A(1). & Tasks 6. a, 6. c, 6.d A(1), Task 5 A(1), Task 4. a A(1), Tasks 3. a, 3. b, 3. c A(1), & Make menu labels specific and match current terminology e.g., rename 'filter' to a more general term such as 'Preprocessing technique' and output prediction to and 'output' to 'export' or 'save.'                                                                                                                                                \\
 \textbf{Issue 3: Inconsistent terminology with the common name for the Task} In multiple instances, the terminology used for tasks, although technically correct, it is not the common term in ML.(e.g., 'partition' for splitting data for train and test, 'one-to many' or 'nominal to binary' for one-hot encoding) &  Task 3. a A(2) & Task 3. a,3.c A(2) & Make sure the tasks' names match current terminology and jargon 
 
 \\
 
 \textbf{Issue 4: The user cannot associate what is an acceptable format to enter to select the column to apply the transformation on }  It is clear that some identifier should be entered in the 'attributeIndices' text field, but as the system had only first-last as a hint, it is not clear how to specify indices in the middle & Task 3.a A(3) & - & Provide a drop-down list with all possible attributes names that the user can apply the filter on them instead of the text box "attributesIndices".
 
 \\
 
 \textbf{Issue 5: Missing label of selecting target drop-down menu} With an overcrowded 'Classify' tab, the drop-down menu from which the target variable can go unnoticed. & Task 3.a A(3) & - & Provide a label for "Select the targeted attributes" or a strong visual that indicates this is a drop-down list. \\
\midrule
   \multicolumn{4}{l}{Issues related to Q4:  If the correct action is performed, will the user see that progress is being made toward the solution of the Task?} \\
    \midrule
 \textbf{Issue 1: lack of feedback after applying transformation}  In Weka, when the user applies the one-hot encoding, save a model or a file, no feedback is provided. Lack of feedback is also observed when trying to visualize data in KNIME. & Task 3.a A(4), Tasks 6.a, 6.d A(4) & Task 2 A(3) & Provide clear feedback of what has been done after each Task immediately. \\

 \textbf{Issue 2: lack of feedback on which connections to use for training or testing} After choosing the train test split, it is hard to tell which connection is to use for training or testing. &-& Task 3.c A(4) & Provide labels on the connections. \\
                                                                                 \bottomrule
\end{tabular}%
}
\end{table*}

\begin{figure*}
  \centering
  \includegraphics[width=0.9\linewidth]{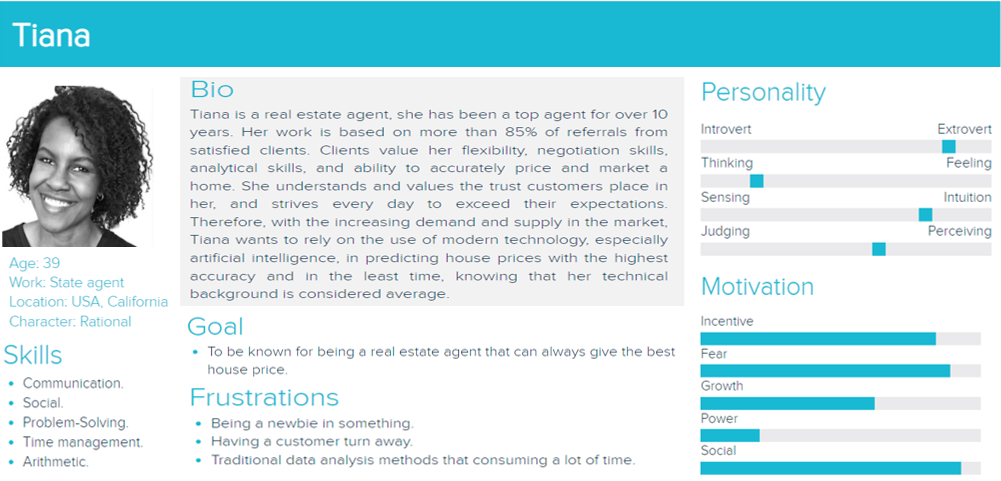}
  \caption{Persona for the user analysis  \label{persona}}
\end{figure*}

\begin{figure}
  \centering
  \includegraphics[width=0.9\linewidth]{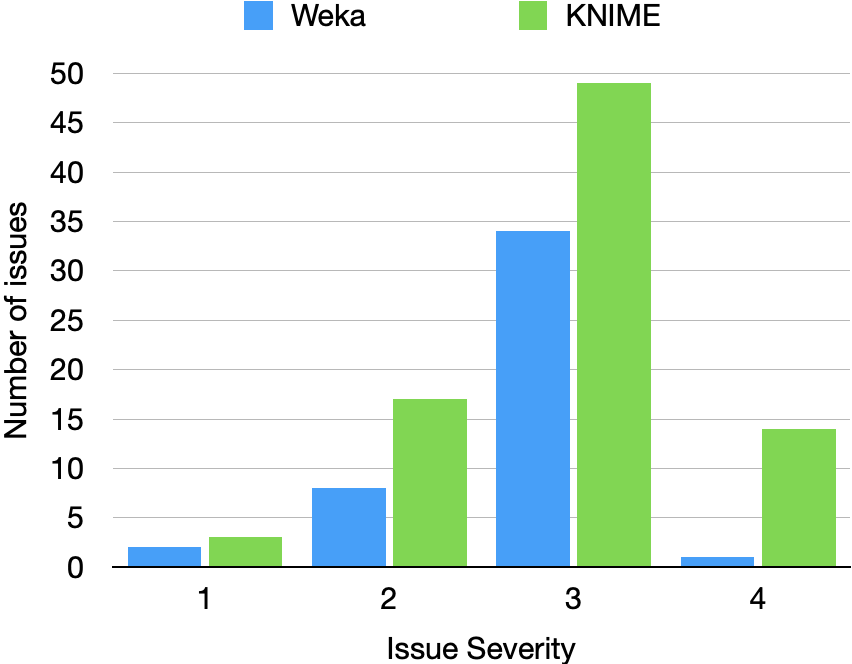}
  \caption{Issues discovered during the Heuristic Evaluation by severity.  \label{fig:sev}}
  \Description[Issues discovered during the Heuristic Evaluation by severity.]{A bar chart displaying the severity of issues discovered during heuristic evaluation on WEKA and KINME ML tools. WEKA has 2,8,34, and 1 issues of severity 1,2,3 and 4, respectively. KNIME has 4,18,49, and 14 of severity 1, 2, 3, and 4, respectively.}

\end{figure}


  

  


\begin{figure}

  \includegraphics[width=0.9\linewidth]{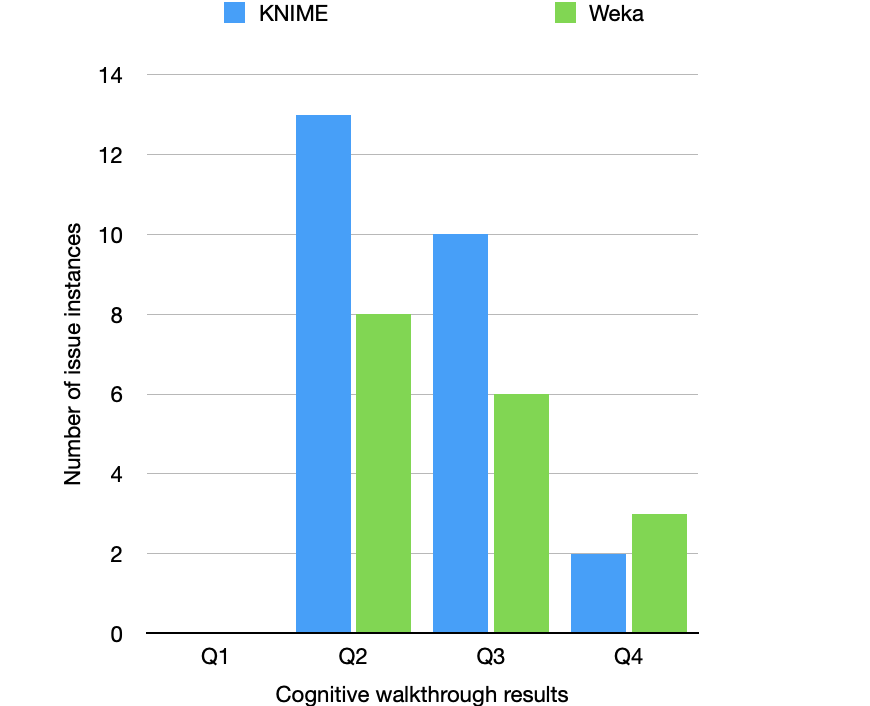}
  \caption{Cognitive walkthrough Results by Question. \label{CWR}}
\Description[Cognitive walkthrough Results by Question.]{A bar chart displaying the total number of issues discovered during the Cognitive walkthrough of WEKA and KINME regarding the four used questions. For WEKA, 0,13,10, and 2 issues for questions 1,2,3 and 4, respectively. For KNIME, 0,8,6, and5 issues for questions 1,2,3, and 4 respectively.}

\end{figure}

\begin{figure}
  \includegraphics[width=0.9\linewidth]{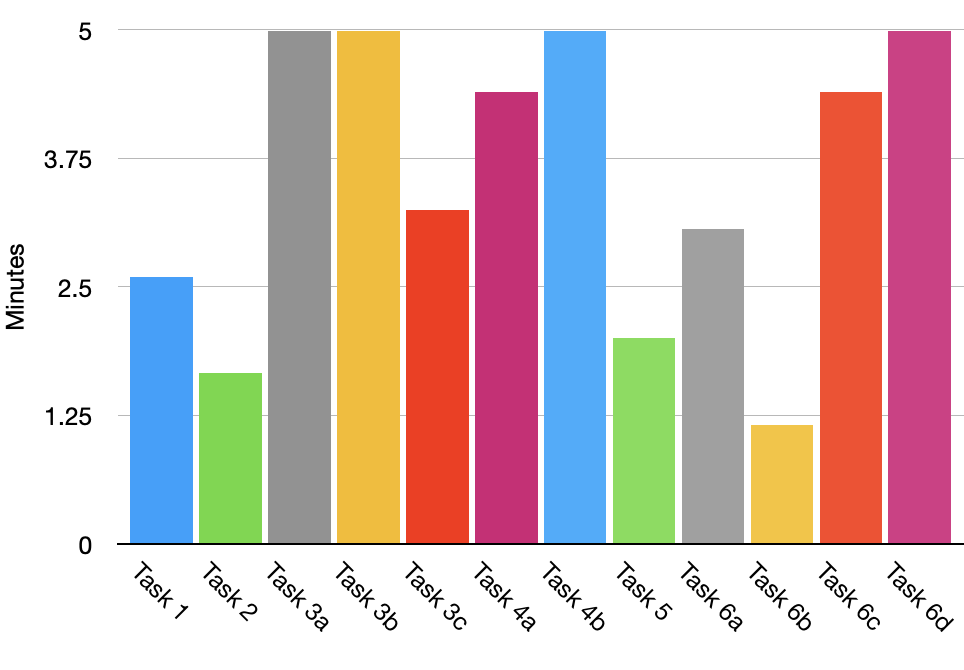}
  \caption{Average Task completion duration (in minutes).}\label{fig:avgp1}
    \Description[Average Task completion duration (in minutes).]{
A bar chart displaying the average Task completion duration (in minutes) of User testing evaluation for WEKA. Tasks 1, 2, 3a, 3b, 3c, 4a, 4b, 5, 6a, 6b, 6c, and 6d have a Task completion duration of 2.60, 1.66, 5.00, 5.00, 3.25, 4.40, 5.00, 2.00, 3.06, 1.16, 4.40, and 5.00, respectively.}

\end{figure}

\begin{figure}
  \centering
  \includegraphics[width=0.9\linewidth]{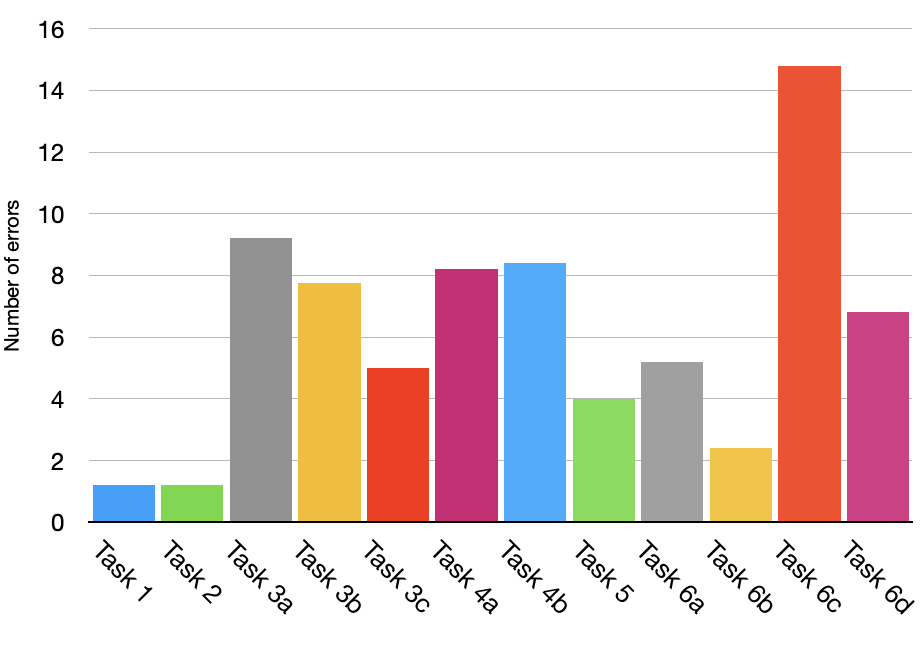}
  \caption{Error rate of Tasks During User Testing of Weka.}
      \Description[Error rate of Tasks During User Testing of Weka.]{
 A bar chart displaying the error rate of tasks during User Testing of Weka. Tasks 1, 2, 3a, 3b, 3c, 4a, 4b, 5, 6a, 6b, 6c, and 6d have an error rate of 1.20, 1.20, 9.20, 7.75, 5.00, 8.20, 8.40, 4.00, 5.20, 2.40, 14.80, and 6.80, respectively.}

  \label{Error-rate-p1}
\end{figure}

\begin{figure}
  \centering
  \includegraphics[width=0.9\linewidth]{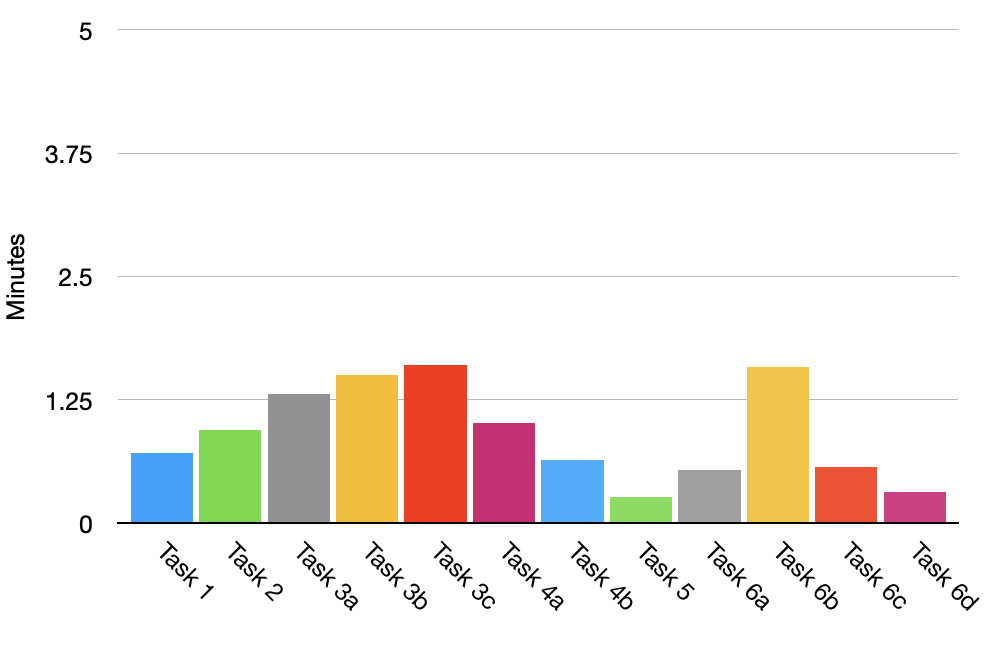}
  \caption{Average Task completion duration (in minutes) - prototype\label{Average-Task-completion-duration-p2}.}
    \Description[Average Task completion duration (in minutes) - prototype.]{A bar chart displaying the average Task completion duration (in minutes) of the user testing evaluation of the prototype. Tasks 1, 2, 3a, 3b, 3c, 4a, 4b, 5, 6a, 6b, 6c, and 6d have a Task completion duration of 0.71, 0.95, 1.31, 1.50, 1.60, 1.02, 0.64, 0.27, 0.54, 1.54, 1.58, 0.57, and 0.32, respectively.}

\end{figure}

\begin{figure}
  \centering
  \includegraphics[width=0.9\linewidth]{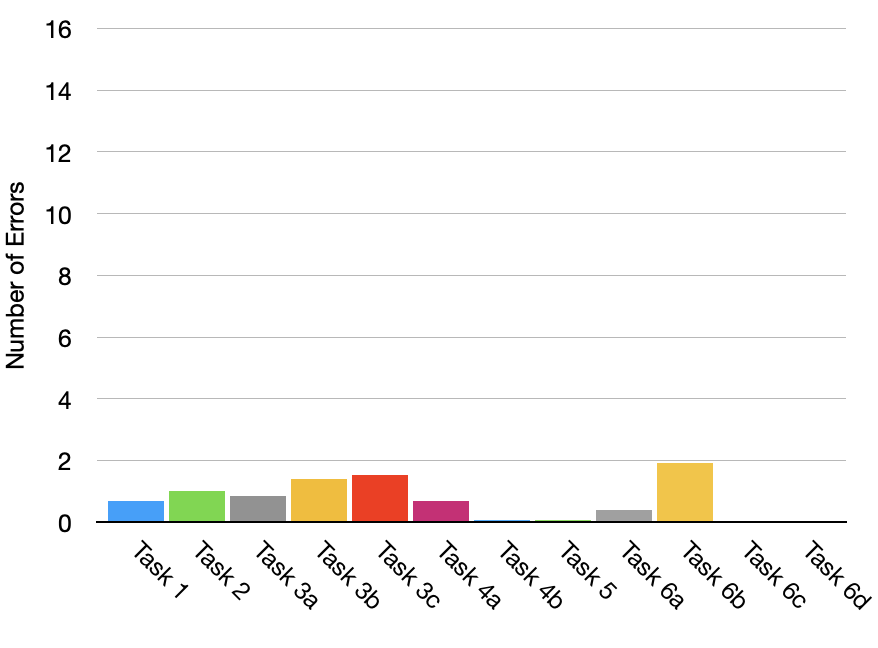}
  \caption{Error rate of Tasks During User Testing of Prototype.}
  \Description[Error rate of Tasks During User Testing of Prototype.]{
A bar chart displaying the error rate of tasks during User Testing of the prototype. Tasks 1, 2, 3a, 3b, 3c, 4a, 4b, 5, 6a, 6b, 6c, and 6d have an error rate of 0.69, 1.00, 0.85, 1.38, 1.54, 0.69, 0.08, 0.08, 0.38, 0.38, 1.92, and 0.00, respectively.}

  \label{Error-rate-p2}
\end{figure}


\begin{figure*}
  \centering
  \includegraphics[width=0.7\textwidth]{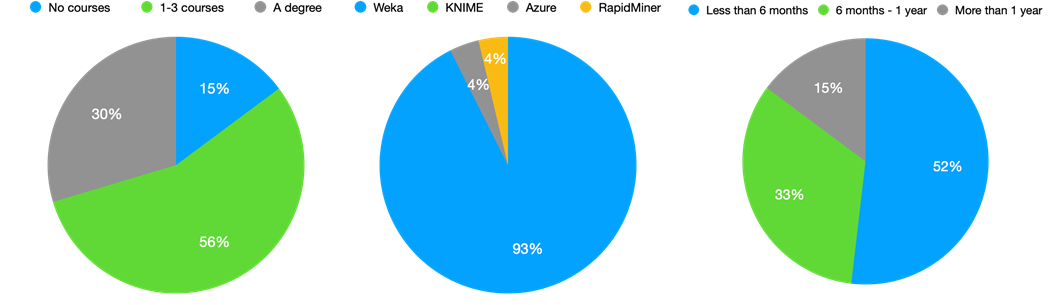}
  \caption{Previous experience with ML tools and educational materials}
  \label{Fig7}
\end{figure*}
\clearpage
\end{document}